**Learning what they think vs. learning what they do:**

**The micro-foundations of vicarious learning**

Sanghyun Park
INSEAD
1 Ayer Rajah Avenue, Singapore
sanghyun.park@insead.edu

Phanish Puranam
INSEAD
1 Ayer Rajah Avenue, Singapore
Phanish.Puranam@insead.edu

**ABSTRACT**

Vicarious learning is a vital component of organizational learning. We theorize and model two fundamental processes underlying vicarious learning: observation of actions (learning what they do) vs. belief sharing (learning what they think). The analysis of our model points to three key insights. First, vicarious learning through either process is beneficial even when no agent in a system of vicarious learners begins with a knowledge advantage. Second, vicarious learning through belief sharing is not universally better than mutual observation of actions and outcomes. Specifically, enabling mutual observability of actions and outcomes is superior to sharing of beliefs when the task environment features few alternatives with large differences in their value and there are no time pressures. Third, symmetry in vicarious learning in fact adversely affects belief sharing but improves observational learning. All three results are shown to be the consequence of how vicarious learning affects self-confirming biased beliefs.

Learning from the experience of others, a.k.a "vicarious learning" is an important component of organizational learning. It can occur at different levels of aggregation (e.g. individuals, groups, and companies) and between as well as within organizations. Vicarious learning can serve as an important source of organizational advantage by improving the utilization of experience within and between organizations (Argote, 2012; Szulanski, 1996). However, despite its centrality to organizational learning, the micro-foundations of vicarious learning have remained largely under-theorized (Bresman, 2010, p. 93; Darr, Argote, and Epple, 1995, p. 1761).

Vicarious learning is known to potentially unfold through two different processes – learning from observing the experiences of others (i.e. "observational learning") (e.g. Gaba and Terlaak, 2013) and learning the beliefs that others have formed based on their own experiences (i.e. "belief exchange" or "knowledge transfer", (e.g. Argote, 2012, Chapter 6)). While the distinction between these processes is intuitive, the implications of this distinction are yet to be theoretically developed. For instance, observational learning often presumes a knowledge differential (Levin and Cross, 2004; Osterloh and Frey, 2000). Does this mean that it is not useful in the absence of such a differential? Would the answer differ in the case of belief sharing? Myers (2018) proposes that bidirectional belief sharing between a learner and a model through "coactive vicarious learning" offers several benefits in terms of organizational learning. Yet it is unclear if it is the belief sharing relative to observational learning that is advantageous, or the symmetry. Well-known pathologies such as "groupthink" curtail organizational learning by reducing diversity of beliefs (Janis, 1972), and symmetry can also arise in observational learning.

A deeper understanding of the micro-foundations of vicarious learning is, in general, useful for scholars interested in organizational learning as well as practitioners of knowledge management in organizations (Argote, 2015). Considerations of how to improve vicarious learning also potentially relate to contemporary challenges around how to organize remote working and distributed teams (Ortiz de Guinea, Webster, and Staples, 2012), and collaboration between humans and algorithms (Agrawal, Gans, and Goldfarb, 2018). Accordingly, our goal is to model the two processes of vicarious learning – belief sharing and observational learning – explicitly at the dyadic level to understand them better, and then extend the analysis to larger systems. We draw on well-established models of individual associative (reinforcement) learning to build a model of vicarious learning in dyads (see Heyes, 2018, Chapter 3; Puranam, Stieglitz, Osman, and Pillutla, 2015). Uniquely, we are able to compare models of observation vs. belief sharing in the same task environments with the same interacting agents, and thus study the relative and combined performance of a dyad using either form of vicarious learning.

Our central finding is that both processes of vicarious learning – belief sharing (i.e. learning "what they think") and observational learning (i.e. learning "what they do") can be useful even in the absence of any *ex-ante* knowledge differential among learners. However, there are also important differences between these processes in terms of effective vicarious learning. In particular, observational learning can outperform belief sharing when the task environment is such that there are relatively few alternatives to search, there are large differences in payoffs associated with alternatives, and there is no time pressure. These results arise primarily from one mechanism: both observational learning and belief sharing reduces the impact of *self-confirming biased beliefs* on learning but to different degrees.

Self-confirming biased beliefs refer to a form of inaccuracy in beliefs that is self-perpetuating because acting according to these beliefs prevents their falsification (e.g. Denrell and March, 2001; Battigalli, Francetich, Lanzani, and Marinacci, 2019). We also explain how self-confirming biased beliefs are related to but distinct from self-confirming equilibria (Fudenberg and Levine, 1993; Battigalli, Cerreia-Vioglio, Maccheroni, and Marinacci, 2015), confirmation bias (Nickerson, 1998), sticking points (Rivkin and Siggelkow, 2002), and "hot-stove" effects (Denrell and March, 2001).

The intuition for the mechanism can be stated as follows: vicarious learning improves performance over individual isolated learning even without an initial knowledge differential between learners because it produces information on counterfactuals for the individual learner. Moreover, this benefit of vicarious learning exists even when the observation is incomplete (i.e., observing either action or outcome only) as long as learners can infer counterfactuals. However, because actions across agents become more strongly correlated through belief sharing than observational learning, this also causes a narrowing of the scope of attempted actions, and therefore of the range of counterfactuals on which information is generated. This results in a relative disadvantage of belief sharing compared to observational learning when the task environment is relatively small, payoffs vary widely across actions and there are no time constraints. A surprising result is that the advantage of belief sharing when it exists comes not from the superior insight of any individual learner ex-ante, but from the way the system of vicarious learners curtails the scope of the search, thereby minimizing the opportunity cost of exploration, albeit at the risk of being subject to self-confirming biased beliefs.

The implications for managerial interventions (including the design of technologies) that support vicarious learning are that they must be contingent on the task environment. In human-algorithm collaborative decision making, for instance, investing in "explainability" of an algorithm's reasons for a decision to humans (Lipton, 2018) may not always be beneficial in terms of combined learning outcomes (though there may well be independent regulatory or ethical reasons for it), and distributed work may sometimes be better supported by technologies that allow a fairly thin slice of mutual observability (e.g. Cramton, 2001) rather than rich interactions that support belief sharing (Kiesler and Cummings, 2002). The properties of the task environment analysed in the model point to the contingencies under which these propositions hold.

**PRIOR LITERATURE**

Vicarious learning involves learning from the experiences of others (Argote and Ingram, 2000; Cyert and March, 1963). It provides a surrogate source of inputs to learning, without forcing the focal agent to directly experience the successes and failures that are the normal grist to the mill of experiential learning (Bresman, Birkinshaw, and Nobel, 1999). By affecting the economics of the utilization of experience as a scarce resource, vicarious learning by individuals within organizations as well as between organizations can affect organizational performance (Argote, 2012; Posen and Chen, 2013; Szulanski, 1996). This is routinely reflected in organizational phenomena that range from the micro (e.g. mentoring of new hires – Taylor, Russ-Eft, and Chan, 2005) to the macro (competitive intelligence that firms derive by studying each other's patents, market expansions, and product launches – Greve, 1998, 2000).

Different forms of vicarious learning have been studied in considerable detail by scholars from different backgrounds. For instance, imitating other's practice has been viewed as a key driver of the diffusion of practices (e.g., DiMaggio and Powell, 1983) as well as competitive dynamics between firms (e.g., Lippman and Rumelt, 1982; Posen and Chen, 2013). On the other hand, the studies on best practice or knowledge transfer have emphasized that an organization extrapolates its past (successful) experience by enabling its members to learn from each other's experience (e.g., Szulanski, 1996), which shares the same principle of "copying the successful" that underlies models of cumulative cultural evolution (e.g., Henrich, 2017; Heyes, 2018). Lastly, co-active learning emphasizes the active role of the model in building a shared understanding of its own experience with learners (Grecu and Becker, 1998; Myers, 2018). All are anchored in the basic notion of leaning from the experience of others.

The theoretical foundations of vicarious learning lie in Albert Bandura's work on social learning among children (Bandura, 1977). Subsequently, organizational scientists have developed an extensive literature on learning in organizational contexts that involves a focal actor's observation of a "model" in action, or of the transmission of the rationale that underlies these actions, either in tangible documentation or through social interaction (Myers, 2018). Yet, these two mechanisms of vicarious learning involve fundamental differences that have been largely unrecognized. Learning from observing the actions and outcomes of another requires that these be observable. While these conditions may have been easily satisfied in work contexts that involved physical and collocated effort, it seems far from obvious that they can be met for the virtual knowledge work that characterizes large parts of the economy today (Hinds and Bailey, 2003). Exactly what is observable about the role model may vary qualitatively across contexts: actions, beliefs, or outcomes.

Access to the beliefs of others, either symbolically coded in manuals or through social interactions produces its own challenges. To successfully share knowledge requires more than a transmission of information (Axley, 1984); a substantial foundation of "common ground" – shared concepts and terminology – is necessary for any meaningful semantic exchange to occur (Bechky, 2003; Clark, 1996; Szulanski, 1996). Further, it may require two-sided interaction or "coactive learning", casting the model in a more active role than as a passive exemplar (Myers, 2018). Thus while observability of the other's behaviour is critical for observational learning, access to the rationale and thought process of the other – either in the form of social interaction and communication or through access to documents and artifacts – is essential for belief sharing. Neither is trivial to ensure.

Finally, even if both processes – observation and belief exchange – are feasible, they are not necessarily equivalent in their consequences for vicarious learners. Individuals act on their beliefs. Therefore, observing only the actions of another necessarily conveys less information about the other, as the actions not taken, and their reasons for not taking them, remain hidden to the observer. On the other hand, the ability to share the beliefs of others certainly conveys more information but not necessarily knowledge, if those beliefs are erroneous. Therefore, far from being equivalent, the two processes of vicarious learning may be profoundly distinct in their consequences for the learner, as a function of (as yet) untheorized contingencies.

Tackling these questions requires us to conceptualize the intricate interactions and dynamics of learning involving multiple agents. Fortunately, a well-developed body of work now exists in the organizational sciences on modelling individual and group learning processes (Lounamaa and March, 1987; Lave and March, 1993; Denrell and March, 2001; Posen and Levinthal, 2012; see Puranam et al., 2015 for a review). However, while prior models either feature observational learning (e.g. March, 1991; Lazer and Friedman, 2007; Fang, Lee, and Schilling, 2010) or belief sharing (e.g. Knudsen and Srikanth, 2014; Puranam and Swamy, 2015), no explicit comparison of the two has been conducted so far.

In the next section, we draw on this literature to formulate a formal model of vicarious learning that allows us to compare the relative efficacy of the two mechanisms – beliefs sharing and observation – across contextually important conditions.

**MODEL STRUCTURE**

The structure of our models of vicarious learning can be understood by the following analogy. Imagine two HR managers in a company who are in charge of hiring employees. There are *m* types/profiles of candidates. The relative quality of profiles is unknown ex-ante. Thus, each manager iteratively chooses among the *m* alternatives based on his or her own belief, and they simultaneously update their beliefs based on the information they later obtained on the performance of the candidates they selected. Each manager makes a choice independently since they are in different divisions. However, they can learn vicariously either by observing each other's experiences (for instance, a centralized database that allows each manager to observe the profiles and performance of past hires made by the other) or by sharing beliefs (a platform for collaborative discussion, exchange of documents, notes, etc on the understanding they have developed about each employee profile). In other words, in our model, each manager learns about candidates from two different sources of information – his own and his colleague's experience (Figure 1).

We can also interpret Figure 1 as a case of collaboration between a human manager and a hiring algorithm (Shrestha, Ben-Menahem, and von Krogh, 2019). For the algorithm, observability of past decisions implies that it's training data includes the actions and outcomes of the human manager. Belief sharing could involve introducing rules and constraints on the algorithms based on the manager's understanding. Conversely, the manager may learn from observing the actions and outcomes of the algorithm, or by trying to understand the reasons for the algorithm's decisions using explainable AI technologies.

We now translate the above analogy into a reinforcement learning model on a multi-armed bandit task where each agent chooses an alternative based on his or her own beliefs and then revises them based on the feedback for the chosen actions. In addition, each agent in our model faces an identical task environment – they are learning in the same context, a precondition for vicarious learning. Thus, our model consists of three key components: the task environment, each agent's task representation, and the learning processes.

[Insert Figure 1 about here]

**Task Environment**

In our model, the task environment is described as $m$ alternatives (i.e., $m$ types of candidates) with different (initially unknown to the agents) expected payoffs (i.e., actual quality levels), which are denoted by $\mathbf{\Pi} = \{\Pi_1, \cdots, \Pi_m\}$. In particular, we assume that the task environment is identical across agents in the system to allow vicarious learning. At period $t$, the realized payoff for agent $i$ who chose alternative $j$ ($\pi_{i,j,t}$) is determined by the sum of the expected payoff ($\Pi_j$) and random noise ($u_{i,j,t}$), drawn from the uniform distribution in $(-\varepsilon, \varepsilon)$. By incorporating random noise with parameter $\varepsilon$, we capture the uncertain nature of the task environment. Note that agents may receive different payoffs even when they choose the same action if there exists uncertainty (i.e., $\epsilon > 0$). In the sampling procedure, we randomize the expected payoffs to offset the effect of payoff distribution, but we fix the value of maximum payoff at $\Pi_{max}$ to make the performance of the system comparable across samples. To be specific, among $m$ alternatives, only one will be assigned the value of $\Pi_{max}$, whereas payoffs for other alternatives will be drawn from the uniform distribution in $(0, \alpha)$ where $\alpha > 0$. In other words, $\alpha$ and $\Pi_{max}$ determine the relative attractiveness of the best alternative compared to other alternatives (i.e. "spikiness" of the payoff surface – or by how much the global peak exceeds the next largest local peak).[1]

**Agent's Task Representation and Choice Rule**

Although the task environment has some objective expected values of alternatives, these are unknown ex-ante to both agents so that they make choices based on their own representations, denoted by $\boldsymbol{R} = \{\boldsymbol{R}_{1,t}, \boldsymbol{R}_{2,t}\}$. Each agent's representation consists of his or her own belief on the expected payoff for each alternative (i.e., $\boldsymbol{R}_{i,t} = \{r_{i,j,t}\}$ where $i \in \{1, 2\}$, $j \in \{1, \cdots, m\}$, and $t \in \{0, \cdots, T\}$). At the initial stage (i.e., $t = 0$), each agent possesses own beliefs (i.e., "priors") on a payoff for each alternative that may arise from past experience. In particular, priors will be randomly drawn from the uniform distribution (0, 1). Over time, agents update their representations through learning processes, which will be described in the following section. In this regard, the agent's representation reflects his or her belief about the task environment shaped by both prior and learning processes.

Based on the above conceptualization of representation, we assume that the probability of choosing any alternative depends on the agent's expected payoff for that alternative relative to those for other alternatives, following the softmax rule (Sutton and Barto, 2018). To be specific, the probability that agent $i$ chooses an action $j$ in period $t$ is given by:

$$p_{i,j,t} = \frac{e^{r_{i,j,t}/\tau}}{\sum_{j=1}^{m} e^{r_{i,j,t}/\tau}}$$

where $r_{i,j,t}$ is agent $i$'s belief on the expected payoff for alternative $j$ at period $t$. The parameter $\tau$ reflects the degree of exploration in the search process (Sutton and Barto, 2018).

---

[1] Note that this is not identical to "ruggedness".

When $\tau$ is high, the selection of choices depends less on the subjective valuation of alternatives (i.e., more exploration). On the other hand, as $\tau$ decreases, the agent is more likely to choose the alternative believed to offer higher payoff (i.e., less exploration). In particular, when $\tau \to 0$, the agent is more likely to choose the alternative believed to have the highest expected payoff.

**Learning Processes**

Although priors held by the agents shape their choice in the early stage, they also learn from the feedback on choices they made in the past by modifying their representations in response to feedback. More importantly, in our model, they learn from their collaborators as well (i.e., vicarious learning). In particular, we examine two distinctive processes of vicarious learning: belief sharing and observational learning. Our model thus has three distinctive learning processes – (1) learning from own experience, (2) vicarious learning through belief exchange, and (3) vicarious learning through observation.

*Learning from own experience*

When the agents learn from their own experience, we assume that they engage in reinforcement learning, which has been widely adopted to describe the adaptive behaviour of individuals and organizations (e.g., Herriott, Levinthal, and March, 1985; Lave and March, 1993; Denrell and March, 2001). To be specific, the updating rule for agent $i$ who chose the alternative $j$ and received $\pi_{i,j,t}$ at period $t$ is given by:

$$r_{i,j,t+1} = r_{i,j,t} + \phi(\pi_{i,j,t} - r_{i,j,t}) \qquad (1)$$

where $\pi_{j,t}$ is the realized payoff for the alternative $j$ at period $t$ and $r_{i,j,t+1}$ is the updated belief held by agent $i$ at period $t+1$. The control parameter in reinforcement learning is $\phi \in [0,1]$ that reflects the learning rate in the updating procedure. When $\phi$ is high (i.e., fast learner), the agent is highly sensitive to feedback, thereby rapidly adjusting own belief ($r_{i,j,t}$) toward the realized payoff ($\pi_{i,j,t}$). On the other hand, when $\phi$ is low (i.e., slow learner), the agent adjusts his own belief slowly.

It is important to note the key characteristics of the above model. First of all, there exists the tendency to repeat actions that provided higher payoffs, whereas actions that delivered lower payoffs become less likely to be chosen again (i.e., Thorndike's Law of Effect). Second, the agents assess and update the beliefs by comparing the realized payoffs ($\pi_{i,j,t}$) and the existing belief ($r_{i,j,t}$). Third, the beliefs on all unchosen alternatives remain unchanged in the following period. This is a widely used approach to modelling human reinforcement learning because it embodies three robust psychological signatures of learning: law of effect, law of recency, and the power law of practice (Erev and Roth, 1998).

### *Learning through belief exchange*

The agents learn not only from their own experience but also from others' experience or beliefs (i.e., vicarious learning). The most direct form of vicarious learning is 'belief exchange' where the agents share their beliefs through direct interaction (e.g., discussion). In other words, the agents can learn from others by accessing their belief systems. Given that the agents' task representations in our model reflect their belief, we characterize belief exchange as a function of task representations held by focal and target actors. Specifically, the updating rule for agent 1 who learns from agent 2 by accessing his or her belief system at period $t$ is given by:

$$\boldsymbol{R_{1,t+1}} = (1 - \phi_{BS})\boldsymbol{R_{1,t}} + \phi_{BS}\boldsymbol{R_{2,t}} \qquad (2)$$

Note that the above equation is simply the weighted average of the two belief systems (e.g., Friedkin, Proskurnikov, Mei, and Bullo, 2019). When the learning rate for belief sharing $\phi_{BS} = 0$, the process reduces to individual learning.

*Learning through observation*

Even when the agent cannot access the other's beliefs, he or she can learn from others by observing their actions and/or corresponding outcomes. When the agent observes both action and outcome, the information set obtained through observation can be treated in the same way as that obtained from own experience. Specifically, we assume that observational learning also follows the reinforcement learning. Specifically, the updating rule for agent $i$ who observed that the other agent $i'$ chose the alternative $j$ and received $\pi_{i,j,t}$ at period $t$ is given by:

$$r_{i,j,t+1} = r_{i,j,t} + \phi_{OL}(\pi_{i',j,t} - r_{i,j,t}) \text{ where } i \neq i' \quad (3)$$

where $\phi_{OL}$ is a parameter for the observational learning rate. Note that the learning rate for observational learning ($\phi_{OL}$) can differ from that for direct experience ($\phi$). When two agents choose the same action, we assume that they update corresponding belief twice with two realized payoffs. As there is no agent-specific effect, the order of update does not have any systematic effect on outcomes.

*Learning through incomplete observation*

Although we assume that the agent can observe both other's actions and outcomes in the baseline model, observation may be limited to either action or outcome (i.e., incomplete observation). However, the agents can infer the task environment or other's belief system even when the observation is incomplete. For example, when the agents observe that the other agent received a higher payoff then they did, they will believe that there exists a more attractive alternative. In turn, they are more likely to engage in exploration (i.e., inspiration-based search). In other words, the agent can deduce the distribution of payoffs in the task environment by observing other's outcomes. To capture this process, we assume that the agent will adjust his exploration rate by comparing the other's outcome with his own beliefs. To be specific, the exploration rate (i.e., $\tau$) for the agent $i$ at period $t$ who observed that the other agent received $\pi_{j,t-1}$ at period $t-1$ is given by:

$$\tau_{i,t} = \begin{cases} \tau_L & \pi_{j,t-1} \leq cr^*_{i,t-1} \\ \tau_H & \pi_{j,t-1} > cr^*_{i,t-1} \end{cases} \quad (1)$$

where $r^*_{i,t-1} = \max\{\boldsymbol{R_{i,t-1}}\}$ and $c$ is a constant for the threshold. Note that, in this case, the other's action (i.e., $j$) is unobservable. Put it simply, the agent will engage in more exploration ($\tau_H$) when it observes the other's payoff ($\pi_{j,t-1}$) much higher than the highest payoff in its own beliefs ($cr^*_{i,t-1}$).

Similarly, the other agent's behaviour plays as a cue about his or her belief system since the alternative that is believed to provide higher payoff is more likely to be chosen. When the agent observes the other's action, the agent will consider it as an indirect indication of the attractiveness of that alternative, which, in turn, increases a propensity to try that alternative (i.e., imitation). Here, we assume that the agent adjusts own belief regarding the observed alternative with respect to the maximum expected return ($\Pi_{max}$). Specifically, the updating rule for the agent $i$ who observed that the other agent chose the alternative $j$ at period $t$ is given by:

$$r_{i,j,t+1} = r_{i,j,t} + \phi_{OL}(\Pi_{max} - r_{i,j,t}) \quad (2)$$

Simply put, when the agent observes other's behaviour (i.e., alternative $j$), it will consider that alternative as more attractive (i.e., increases $r_{i,j,t}$). Overall, Table 1 describes what ego can observe about alter with respect to different micro-processes of vicarious learning.

[Insert Table 1 about here]

**RESULTS**

We examine (1) whether vicarious learning can improve performance even without ex-ante knowledge differential and (2) the conditions under which two baseline models of vicarious learning (i.e., belief sharing and observational learning) may produce a relatively better or worse performance at the system level (i.e., dyad). We assess performance as the average payoffs over time of the two agents. The model parameters are described in Table 2. All repeated simulations involve 100,000 runs.

[Insert Table 2 about here]

**Baseline Model: A System with Vicarious Learning between Two Symmetric Agents**

As a baseline case, we first start with the analysis of the learning system with two baseline models of vicarious learning: observational learning (i.e., complete observation) and belief sharing. In particular, we assume that the system consists of two symmetric agents. By symmetric, we mean that the agents in the system exhibit identical learning behaviours (i.e., identical learning rates and exploration rates). Further, we assume 'symmetric influence' where the agents build a shared belief system by assigning identical weights to their own and other agents beliefs (i.e., $\phi_{BS,1} = \phi_{BS,2} = 0.5$). This captures coactive learning as described by Myers (2018) – which involves bidirectional belief sharing between learner and a role model. Finally, we assume that two parameters for learning from own and observation of other's experiences are identical within the agent (i.e., $\phi_{OL} = \phi$). Figure 2 shows the learning performance of the dyad in terms of average payoffs over time for different modes of vicarious learning (as well as without vicarious learning), for model settings of $m = 50$, $\varepsilon = 0.1$, $\phi = 0.5$, $\tau \to 0$). This is a particular point in the parameter space we use for illustration, and later we broaden our search.

[Insert Figure 2 about here]

Compared to the system without vicarious learning, we find that both modes of vicarious learning improve the performance of the system in initial periods. Information regarding other's experiences enhances performance even when the agent does not have access to other's belief systems. Further, observational learning always outperforms individual learning, whereas belief sharing only initially dominates individual learning. Interestingly, we find that allowing both observation and belief sharing (i.e., hybrid) is not necessarily superior to limiting interaction to observation. Rather, our results show that the system with hybrid vicarious learning behaves similarly to that with belief sharing.

Before we investigate the mechanisms underlying these results, it is worth reiterating that neither agent begins with better knowledge, and on average they are equally ignorant of the task environment. They are also symmetric in terms of their learning parameters and only differ in their priors (which are no more accurate than chance – note that performance begins at the same level in all forms of learning in Figure 2). The results thus demonstrate the benefits of vicarious learning even in the absence of superior wisdom ex-ante in the system (which is presumed in phenomena such as best practice transfer). We next turn to the reasons why vicarious learning produces such an advantage relative to individual learning even in the absence of ex-ante knowledge differentials as well as why belief sharing seems to suffer a disadvantage over time but observational learning does not.

Broadly, there are two benefits of learning from the experience of others. The first is a fairly mundane effect of increasing sample size on cancelling out the noise in observation (i.e. sampling error). If in a single period two agents pick the same choice and are able to observe each other's actions and outcomes, then their estimate of the expected payoff of that action necessarily becomes more precise. Vicarious learning thus serves as a source of "increasing sample size" by allowing experiences to be aggregated across agents, simply speeding up the learning process. However, this "increasing sample size" (or statistical efficiency of the learning process) is not the only driver of the superior performance of vicarious learning.

A second and more subtle effect arises because both observational learning and belief sharing can affect the impact of *self-confirming biased beliefs* (SCBB). There are two conditions under which SCBB can arise: (1) information gathered (in the form of outcomes) is contingent on the action chosen (i.e., own-action dependence), and (2) actions also depend on information obtained (e.g. rationality as consistency) (Figure 1). Under these conditions, the agent may construct beliefs that are self-perpetuating because acting on these beliefs leads to confirming the existing beliefs, which in turn shapes the subsequent action. When self-confirming beliefs are biased, incorrect beliefs will self-perpetuate without being falsified (i.e., self-confirming biased beliefs). For instance, consider an agent who believes that alternatives A and B yield 50 and 100 units of payoff, when in reality they are worth 150 and 100 (biased beliefs). If the agents take actions consistent with their beliefs, they should select alternative B. The resulting outcome will be 100, increasing their confidence in their beliefs and making the selection of alternative A even less likely in the future (self-confirming the biased beliefs).

In *n*-agent games, SCBB may result in self-confirming equilibria, which may deviate from Nash equilibria (Fudenberg and Levine, 1993; Battigalli et al., 2015). When there is no interdependence between agents, there are two sources of SCBB's: erroneous priors and noisy feedback. For example, when the agent believes that a specific alternative is unattractive (i.e., negative belief), such belief will not be revised even when it is incorrect (i.e., a false-negative belief) since that alternative will not be sampled. On the other hand, even when the optimal alternative (i.e., highest expected payoff) has been sampled, the agents may deviate from that alternative in the subsequent periods if the realized payoff is below the expected payoff due to the noise in payoffs (i.e., "hot-stove" effect, Denrell and March, 2001). SCBB's are thus a superset of self-confirming equilibria (because they can exist even with a single agent) as well as "hot stove effects" (because they can exist even when there is no noise in payoffs).

SCBB are distinct from sticking points, which refer to a configuration of choices such that, once the adaptive system arrives at the configuration, it will never deviate from it (Rivkin and Siggelkow, 2002). While SCBB and sticking points share a similarity in the sense that both are attractors (stable fixed points) of the adaptive system, the source of stability varies. On the one hand, sticking points arise due to the interdependency between elements of the system. An accurate assessment of these produces fixation for the agent. On the other hand, SCBB arise due to the combined properties of the agent's mental representation and the task environment rather than interdependency in the task environment alone. Specifically for SCBB, the agents beliefs must be biased in a way that the actions they stimulate prevent the generation of evidence that may disconfirm the belief. SCBB can arise even without interdependency in the task environment (e.g., as in our model).

Also, SCBB differ from confirmation biases. Confirmation bias refers to "*the seeking or interpreting of evidence in ways that are partial to existing beliefs, expectations, or a hypothesis in hand* (Nickerson, 1998)." In other words, it is a cognitive bias in the information processing process (i.e., biased search, biased interpretation, and biased memory) driven by cognitive heuristics and cognitive dissonance (e.g., MacCoun, 1998). However, the root cause for SCBB is a task environment that forces experience-based (or "endogenous") sampling – the evidence a learner gathers is dependent on their action (Denrell, 2005; Sutton and Barto, 2018).

This is because the task environment requires the learner to "sample to earn", not merely "sample to learn"- payoffs to taking actions matter forcing pressures towards consistency of action and belief in Figure 1. Put differently, adaptively rational agents may suffer self-confirming biased beliefs even when they are unbiased in processing information. Indeed, studies have shown SCBB do not depend on the specific learning process. For instance, SCBB's have been documented in models where agents follow psychologically validated learning rules (e.g., Denrell and March, 2001) as well as Bayesian updating (e.g., Battigalli et al., 2019; Le Mens and Denrell, 2011).

There are only two ways to break SCBB. The first involves forcing the agent to take actions inconsistent with their own beliefs. In learning models, this is the exploration process (tuned by $\tau$)[2] or to provide evidence to the agent that does not arise only from the agent's own actions (i.e. supply information on counterfactuals). Vicarious learning instantiates the second solution, as we shall show, but observation and belief learning feature it to different degrees, which produces differences in their outcomes.

With observational vicarious learning, each agent can also gather the information that is exogenous to his or her own action, thereby escaping own experience-based sampling and exclusive dependence on own beliefs. Unless the learners hold identical beliefs across whole learning periods (i.e., identical priors and learning rates, and no noise in payoffs), they will in general take different actions, and observing the outcomes of these actions can help to break self-confirming biased beliefs for both. With belief sharing, each agent updates her beliefs not only about the action she has taken but about the entire set of alternatives, since agents will typically start with different beliefs. Again, this can help break SCBB. In either case, the power of vicarious learning comes not necessarily from the transmission of superior insight from one agent to another, but from diversity in (possibly erroneous) beliefs.

[Insert Figures 3 about here]

---

[2] James March memorably referred to this as a "technology of foolishness" (1976)

Panel A in Figure 3 illustrates these effects by plotting the joint probability for *both learners* to choose the optimal action against time for different learning systems. First of all, our result shows that all three systems in our model are susceptible to SCBB (i.e., the probability of choosing the optimal outcome is less than 1 even when *T* is large). Further, suffering from SCBB does not automatically imply poor absolute performance. Note that being stuck on a sub-optimal action may nonetheless yield high absolute levels of payoff (compare figures 2 and 3A), if the task environment is such that the difference in payoffs between the optimal and other actions is not large (i.e. the payoff landscape is flat, not spiked).

More importantly, we find that the two vicarious learning systems reduce SCBB, measured as a stable tendency to settle on actions other than the optimal actions, but to different degrees (see Appendix A). Even though individual learning does improve in its ability to find optimal actions with increasing $\tau$ (the first path to escaping self-confirming biased beliefs), across all values of $\tau$, vicarious learning outperforms individual learning (indicating it is the second path – exposure to counterfactual information about actions not taken that primarily creates the advantage for either form of vicarious learning). However, belief sharing clearly loses its early advantage in terms of aiding discovery of optimal actions to observational learning. This is because it loses its ability to produce information on a wide range of counterfactual alternatives over time.

To demonstrate this, Panel B in Figure 3 plots the convergence in actions across the agents over time for all three learning systems. Agents converge much faster to the same set of actions under belief sharing (because they develop similar beliefs on the value of all alternatives), which can reinforce self-confirming biased beliefs if the actions they believe are optimal are in fact not the optimal ones. In other words, other's experience acquired through belief sharing is less likely to resolve the problem of endogenous sampling compared to that in observational learning.

Panel C in Figure 3 shows that this pattern of rapid convergence of actions through belief sharing is generally robust to different weights on own and other's beliefs in belief sharing as well as the exploration parameter, τ. These results explain why in Figure 2, we observe that vicarious learning performs better (in terms of average payoff over time) compared to individual learning for initial periods, but also begins to explain why belief sharing eventually underperforms both individual learning and observational learning. This also explains the surprising result in the same figure that the system with hybrid vicarious learning is not superior to observational learning in the long run.

To illustrate that the increase in "sample size" created by vicarious learning alone cannot explain the advantage of vicarious learning, in Figure 4, we show results for all three systems when in each, agents always receive feedback information on all actions in every period (including the ones they did not select). While there is an advantage for vicarious learning in terms of discovering the optimal alternative, it is dissipated within very few periods. The persistent differences we observed in Figure 2 cannot have arisen from the pure statistical efficiency of vicarious learning but requires the bias reduction discussed above.

[Insert Figure 4 about here]

**Incomplete Observation**

In this section, we investigate whether the benefit of vicarious learning without ex-ante knowledge differential still exists even when the observation is incomplete (i.e., only the outcome or action of the peer is observable). Figure 5 illustrates the average cumulative learning performance of the system across different learning rates and modes of vicarious learning. Our results show that incomplete observation can improve the learning performance of the system compared to isolated learners but only in certain circumstances. This is because the agent can infer counterfactuals based on incomplete information from others, and the relative benefit (and cost) of such information depends on the noise in the information generating process. On the one hand, when they observe other's outcomes but not actions, learning performance increases if the uncertainty of the task environment is low (Figure 5a). This is because the agent can infer the payoff structure of the environment based on other's performance. When they observe that other's payoff is much higher than their payoffs, they are more likely to engage in exploration, thereby correcting SCBB (i.e., inspiration-based search). However, when the environment is highly uncertain, the information from others (i.e., other's payoff) may be driven by noise, which incurs the opportunity cost of search without benefit.

On the other hand, when they observe other's actions but not outcomes, learning performance can be improved if the exploration rate is sufficiently low (Figure 5b). This is because others' actions are likely to reflect their beliefs so that the agent can indirectly infer their belief systems by observing their actions. Moreover, as adaptive agents are likely to choose an alternative that is believed to be attractive, this information will increase the attractiveness of that action in a focal agent's belief system (i.e., imitation). This again can reduce SCBB of a focal agent as long as they choose different alternatives. However, when the exploration rate is high, other's actions may not reflect their belief systems, thereby harming the learning performance of the system.

[Insert Figure 5 about here]

**Exploring Contingencies: Variation in the Task Environment**

We now explore the diverse environmental conditions to understand boundary conditions for two baseline models of vicarious learning: observational learning (complete observation) and belief sharing. To assess the relative performance of two collaborative systems, we compare the cumulative learning performances of the system in a given period. In particular, we focus on three features of the task environment: the number of alternatives (i.e., search space, $m$), the payoff distribution (specifically how flat or spiky it is), and learning periods ($T$). We compute the model for all combinations of learning rates among the agents.

*The effect of the number of alternatives*

Figure 6 illustrates how the number of alternatives affects the relative learning performance of the two systems. Here, we also allow the system to have asymmetric learning rates to explore all possible parameter space. Our results show that observational learning outperforms belief sharing when the search space is relatively small (i.e., $m$ is small). For example, when $m$ is 10 (Figure 6a), the optimal performance of observational learning is higher than that of belief sharing. As the search space becomes broader, the probability of finding the best alternative decreases, which in turn reduces the expected return for search. Another interesting feature is that, in observational learning, the learning performance reaches the optimal level when the agents have different learning rates (i.e., asymmetric agents). This is because asymmetric adjustment by agents in observational learning can balance the impacts of prior and uncertainty on SCBB. To be specific, a slow learner (i.e., one that is less sensitive to the most recent outcome) provides stability in sampling, thereby reducing the impact of uncertainty on generating SCBB. On the other hand, a fast learner enables the system to rapidly deviate from false-positive beliefs, thereby reducing the impact of priors on SCBB (Please see Appendix B for details). However, asymmetry is not useful for belief sharing, as beliefs are merged. On the other hand, when the search space is large (i.e., *m* is large), we find that belief sharing performs better than observational learning. By producing rapid convergence, belief sharing avoids fruitless exploration.

[Insert Figure 6 about here]

*The effect of "spikiness" in payoff distribution*

Figure 7 represents the effect of spikiness (the difference between global peak and next best alternative) in payoff distribution on the relative performance of two modes of vicarious learning. In particular, we operationalize spikiness in payoffs with two parameters ($\Pi_{max}$ and α) without loss of generality. The gap between the two parameters represents the extent of spikiness- the relative attractiveness of the best alternative compared to the next highest alternatives. On the one hand, when the gap between two parameters is large (i.e., high spikiness in payoffs), finding the best alternative becomes critical for learning performance. On the other hand, when the gap between two parameters is small (i.e., low spikiness in payoffs), engaging extensive search to find a better alternative might result in an inferior outcome due to the opportunity cost of search with limited benefit. Our results show that belief sharing outperforms observational learning when spikiness in payoffs is relatively low (Figure 7a). However, as spikiness in payoffs increases, observational learning dominates belief exchange in terms of the optimal performance (Figure 7b). The more spiky the payoff surface (i.e. the larger the difference between global peak and other alternative's payoffs), the more the advantage shifts towards observational learning.

[Insert Figure 7 about here]

*The effect of learning period (T)*

Figure 8 represents the effect of learning periods (i.e., *T*) on the learning performance of two collaboration modes. In the adaptive learning process, the learning periods determine optimal learning behaviour. For example, when the agents face limited time for learning, exploring for better alternatives might lead to inferior outcomes since they might not have enough time to exploit those alternatives to compensate for the opportunity cost of exploration. Our results show that belief sharing outperforms observational learning when they encounter limited learning periods (Figure 8a). However, observational learning performs better than belief sharing when there are less time constraints (Figure 8b).

[Insert Figure 8 about here]

**Mechanisms for Boundary Conditions**

In the previous section, we conclude that observational learning dominates belief sharing in redressing SCBB. However, it is important to note that SCBB do not always imply an inferior outcome. To find the optimal outcome, the adaptive agent may need to engage in extensive search, which induces the opportunity cost of exploration. We find that the system with observational learning has a broader search scope (i.e., exploring more alternatives) compared to that with belief sharing (Appendix C). In this regard, the relative advantage of observational learning (or belief sharing) depends on the relative benefit of correcting SCBB compared to the opportunity cost of an extensive search. To be specific, observational learning dominates belief sharing when correcting SCBB is critical for performance, whereas the advantage shifts towards belief sharing as the opportunity cost of correcting SCBB become prominent.

Our analysis of the task environment captured the essential determinants of the relative cost of SCBB compared to the opportunity cost of exploration. For example, as search space increases ($m\uparrow$), the probability of finding the best alternative will decrease, thereby increasing the opportunity cost of correcting self-confirming biased beliefs. On the other hand, when the variability (spikiness) in payoffs increases, the potential gain of correcting SCBB will increase. Lastly, when there are less time constraints ($T\uparrow$), finding the best alternative (i.e., correcting SCBB) becomes attractive since the system can exploit the discovered alternative for a longer period of time. Table 3 provides a summary of our key findings.

[Insert Table 3 about here]

**Robustness and Additional Analysis**

In this section, we explore whether our findings are robust to alternative model specifications. First of all, the updating rule we use might affect SCBB especially when there is noise in payoffs. In particular, simple averaging also has appeal as an instance of the Bayesian norm even though the updating rule in our model (i.e., exponentially recency weighted average or EWA) reflects an empirically derived behavioural regularity (i.e., the law of recency). We found that the pattern of SCBB with respect to modes of vicarious learning is robust to updating rules (i.e., EWA, averaging, or hybrid; see Appendix D). Also, one can argue that biased priors (i.e., $\alpha < 1$) may lead to SCBB or that our findings may be driven solely by noise in payoffs. However, our finding is robust even when prior is unbiased (i.e., $\alpha = 1$) and there is no noise in payoffs. This confirms that SCBB can arise even without noisy feedback. Moreover, unbiased priors, feedback without noise, and Bayesian updating even jointly do not affect our main findings (see Appendix E). This illustrates that SCBB is an umbrella category, which can be instantiated in more specific forms under particular circumstances (e.g. "hot stove" or "self-confirming equilibria").

Second, we have limited our attention to the smallest collaborative system (i.e., two agents) in the previous section. Thus, we examine whether our finding is robust to system size. In particular, we explore learning systems with two different network topologies: random network and 2D lattice with Von Neumann neighborhood. In Appendix F, we show that vicarious learning improves the learning performance of the system even when there is no ex-ante knowledge differential. Moreover, by using one of the contingencies we discovered (i.e., spikiness), we reconfirm that observational learning outperforms belief sharing especially when SCBB is critical. In contrast, belief sharing dominates observational learning when the opportunity cost of correcting SCBB is greater than its benefit. Overall, these results show that our findings at the dyadic level are robust even when we extend them to larger systems.

Lastly, heterogeneity between observational learning and belief sharing may be driven by differences in the amount of information exchange. To be specific, agents with belief sharing in the baseline model exchange all dimensions of representation every period, whereas the agent with observational learning updates chosen actions only. This asymmetry in the amount of information exchange may make belief sharing more susceptible to SCBB compared to observation. To examine these issues, we explore whether incomplete belief sharing undermines SCBB further than observational learning. We find that belief sharing is still more susceptible to SCBB than observational learning even when they share a limited number of dimensions or share less frequently (Appendix G). Interestingly, the results show that sharing beliefs on chosen actions outperforms sharing beliefs on random actions in redressing SCBB. This is because beliefs on chosen actions are less likely to be biased than those on unchosen actions. Overall, these findings imply that the convergence in actions through sharing private beliefs is a major driver of our findings rather than other factors (e.g., the amount of information exchange).

**CONCLUSION**

In this study, we explore two fundamental processes underlying vicarious learning. We show that these differ systematically in the ability to redress self-confirming biased beliefs, which in turn implies that their relative advantage depends on the relative benefit of correcting self-confirming biased beliefs compared to the opportunity cost of exploration. Put simply, sharing belief does not always dominate sharing observation for producing superior performance. Sharing private beliefs leads to the rapid convergence of actions across agents, thereby becoming more susceptible to self-confirming biased beliefs compared to observational learning. On the other hand, when vicarious learning is limited to observation, the other's experience is more likely to remain exogenous to his or her own action, thereby reducing self-confirming more than belief sharing. This tendency provides the relative advantage of observational learning compared to belief sharing when overcoming self-confirming biased beliefs. However, when the benefits of correcting self-confirming biased beliefs become less prominent, belief sharing dominates observational learning.

More broadly, our analysis contributes to theories of organizational learning by helping to sharply demarcate the general process of vicarious learning from its special cases such as imitation (i.e. a form of observational learning (e.g., Gaba and Terlaak, 2013; Posen and Chen, 2013)), best-practice transfer (which could involve either belief exchange or observation but with the model presumed to have superior knowledge ex-ante (e.g., Szulanski, 1996)) and co-active learning (a form of belief sharing (Myers, 2018)). We show that vicarious learning is useful even without ex-ante differentials in wisdom among learners, and also highlight self-confirming biased beliefs as the theoretical mechanism that helps understand the two different processes of vicarious learning – observation and belief sharing- and their relative efficacy.

This study also contributes to the literature on the value of (cognitive) diversity by elucidating its role in a dynamic process of learning. While the diversity literature has highlighted its potential advantage in information processing, which corresponds to the noise cancelling function of vicarious learning, its theoretical framework has largely relied on a static perspective (Srikanth, Harvey, and Peterson, 2016). Filling this gap, our analyses highlight that diversity can be beneficial for a system not only statically but also dynamically. Having colleagues with heterogeneous beliefs in the vicarious learning process provides counterfactuals for individual learners, thereby allowing them to correct self-confirming biased beliefs. This, in turn, will shape their subsequent behaviors in their own learning processes.

Moreover, our result shows that this dynamic benefit of diversity in vicarious learning will be amplified when interactions among team members are somewhat limited (this is in contrast to the emphasis on interactive learning proposed by Myers (2018)). We know that initial team diversity tends to decrease over time as they interact with each other due to group thinking and assimilation process (Harrison, Price, Gavin, and Florey, 2002; Pelled, Eisenhardt, and Xin, 1999). Such a process, we show, may still produce the benefits of vicarious learning through belief sharing, but may underperform observational learning. Put differently, the empirical observation that vicarious learning often engtails rich interaction and belief sharing does not rule out the counterfactual (i.e. that observational learning could have done better), and/or may indicate that the empirical context being observed is one that has the properties that privilege belief sharing (i.e. large space, time limits, low spikiness).

These results not only enhance our understanding of the mechanisms of vicarious learning but also have practical implications. First, given the increasing incidence of remote collaboration (particularly in response to the COVID19 pandemic), the role of technology has become central in bridging the distance not only between actors within an organization (Ortiz de Guinea et al., 2012; Mesmer-Magnus, DeChurch, Jimenez-Rodriguez, Wildman, and Shuffler, 2011) but also between actors across organizations (e.g. online communities, see Faraj, Jarvenpaa, and Majchrzak, 2011; Faraj, von Krogh, Monteiro, and Lakhani, 2016). However, from the perspective of improved vicarious learning, a fundamental design choice for organizing distributed work is between creating technologies that enhance observational learning (for instance electronic repositories and version control software that are extensively used in software development – Cramton, 2001; Gutwin, Penner, and Schneider, 2004), vs. those that enhance belief sharing (e.g. system documentation, text, voice, and video chat – Kiesler and Cummings, 2002). Which should be preferred under what circumstances?

Our results suggest that technologies that enable in-depth interaction are not always superior to those that simply enable mutual observability of actions. This contradicts a common belief underlying the recent effort to enhance the ability to share beliefs through technological development (Metcalf, Askay, and Rosenberg, 2019; Franz, Shrestha, and Paudel, 2020). Our findings underscore that electronic mediated collaboration (e.g. remote working) can be more productive with mere observability of each other's actions and outcomes (Srikanth and Puranam, 2014). New technologies such as Artificial Swarm Intelligence (ASI) enable a group of people to pool private beliefs more extensively compared to traditional methods (e.g., group discussion). For example, ASI allows individuals to share beliefs without requiring explicit or face-to-face communication, thereby overcoming the size limit as well as socio-psychological biases in group discussions (Metcalf et al., 2019). In contrast, algorithmic assistants combined with knowledge repositories can encourage observational learning by providing information regarding successful actions that had been taken by others in similar contexts (Malone, 2018, p. 40). Our results also speak to when the former may be superior to the latter.

Collaborative learning between humans and artificial intelligence (AI) algorithms is another emerging area of interest (Agrawal et al., 2018). What may uniquely distinguish collaborative decision making by humans and AI from other forms of automation or technology use is the mutual adaptation between the two (admittedly different) types of learning systems. Across sectors, organizations have been experimenting for a while with teams of humans and algorithms that learn in the same task environment while also learning vicariously from each other (Grecu and Becker, 1998; see Shrestha et al, 2019 for a recent review). For instance, human and algorithmic fund managers or HR managers might both improve their selection strategies by vicarious learning from each other.

However, a basic challenge with the most sophisticated machine learning algorithms today lies in their explainability. Because they often rely on non-linear function approximation, we are limited as humans in being able to understand why the algorithm does what it does (Lipton, 2018; Samek, Wiegand, and Müller, 2017). This suggests that the channel of vicarious learning through belief sharing may be constrained in such interactions. The key design question again becomes one of choosing between creating technologies that improve the possibility of belief sharing (through explainability) at the cost of predictive power (Lipton, 2018; Gunning, 2017) vs. simply improving the observability of the choices and outcomes that the algorithm (and human) experiences. Again, our results question the imperative to improve the explainability of AI for human-algorithm interaction (e.g., Gunning, 2017). Human-algorithm collaboration can be productive without explainability, under the right conditions.

We recognize that ours is a theoretical exercise and is, in particular, an exercise in normative theory. The assumptions in our theorizing are grounded strongly in the scientific evidence on human behaviour, but the predictions remain to be verified in future research. Specifically, our findings introduce testable hypotheses for contexts where collaborative learning occurs, and these may extend beyond the cases of remote collaboration and human-algorithm interaction we have discussed in detail above. For example, many organizations now engage in search for new ideas through crowdsourcing platforms to stay innovative (e.g., Dahlander and Piezunka, 2014; Gambardella, Raasch, and von Hippel, 2017). To encourage participants to share private beliefs, organizations often interact with them by providing feedback on their ideas (Franke, Keinz, and Klausberger, 2013; Piezunka and Dahlander, 2015). Our findings explain why in-depth interactions between organizations and crowds may result in the premature convergence on ideas submitted. Moreover, feedback will undermine the ability of crowdsourcing to correct biased beliefs of an organization as it reflects such biased beliefs and shapes ideas submitted by crowds. Strategic alliances often also feature collaborative learning processes (see Inkpen and Tsang, 2007 for a review). Our model predicts that allowing close interactions between firms may provide an advantage in the early stage but may result in a premature lock-in to an inferior solution in the long run. Empirical tests of the moderators of this baseline effect (such as the nature of the task environment and time constraints) can provide additional opportunities for falsification of the proposed theory.

# TABLES AND FIGURES

**Table 1. What ego can observe about alter with respect to the mode of vicarious learning**

|  |  | Belief | Action | Outcome |
|---|---|---|---|---|
| | Belief sharing | Yes | | |
| Observational learning | Baseline model (complete observation) | | Yes | Yes |
| | Imitation | | Yes | |
| | Inspiration | | | Yes |

**Table 2. Model parameters for the main results**

| Parameter | Definition | Possible values | Sampled space |
|---|---|---|---|
| $\phi$ | Learning rate | Real number $[0, 1]$ | $[0, 1]$ |
| $\tau$ | Degree of exploration | Real number $(0, \infty)$ | 0.1, 0.01, and $\tau \to 0$ |
| $m$ | The number of alternatives | Positive integer | 10, 50 |
| $\varepsilon$ | Degree of uncertainty | Real number $[0, \infty)$ | High: 1 Low: 0.1 |
| $T$ | Learning period | Positive integer | 1 ~ 1,000 |
| $\Pi_{max}$ | The maximum expected payoff | Real number | 1, 1.5 |
| $\alpha$ | The upper bound for other alternatives | Real number | 0.8 |

**Table 3. A comparison between observational learning and belief exchange**

| | Search scope | Asymmetry in learning rate | Self-confirming biased beliefs | | Boundary condition |
|---|---|---|---|---|---|
| | | | *Due to prior* | *Due to noise in payoffs* | |
| **Observational Learning** | Broad | Increases performance | Less susceptible | Less susceptible (can be redressed by asymmetry) | *When superior*<br>- Narrow search space<br>- High variance in payoffs (spiky)<br>- Sufficient learning period |
| **Belief Exchange** | Narrow | Decreases performance | More susceptible | More susceptible | *When superior*<br>- Broad search space<br>- Low variance in payoffs (flat)<br>- Limited learning period |

**Figure 1. Illustration of vicarious learning process**

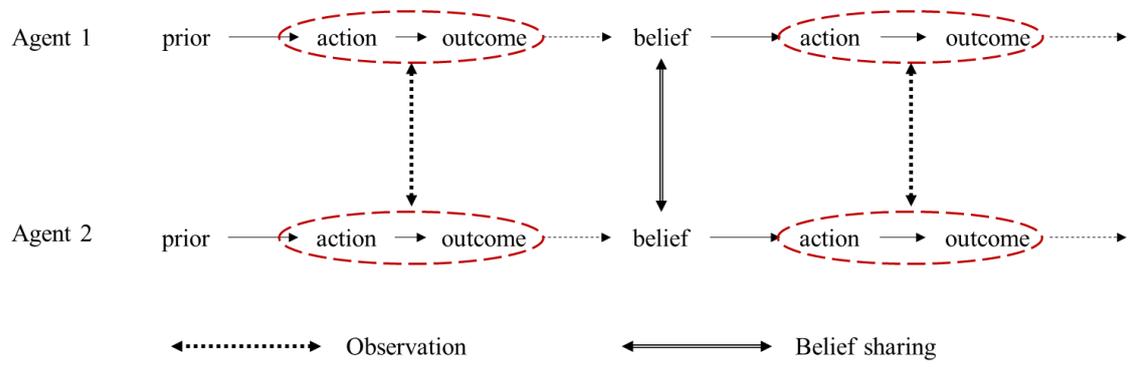

**Figure 2. Intertemporal pattern of learning performance ($\tau \to 0$ ($\max\{R_i\}$), $\phi = 0.5, \Pi_{max} = 1, \alpha = 0.8, \varepsilon = 0.1, T = 1{,}000$)**

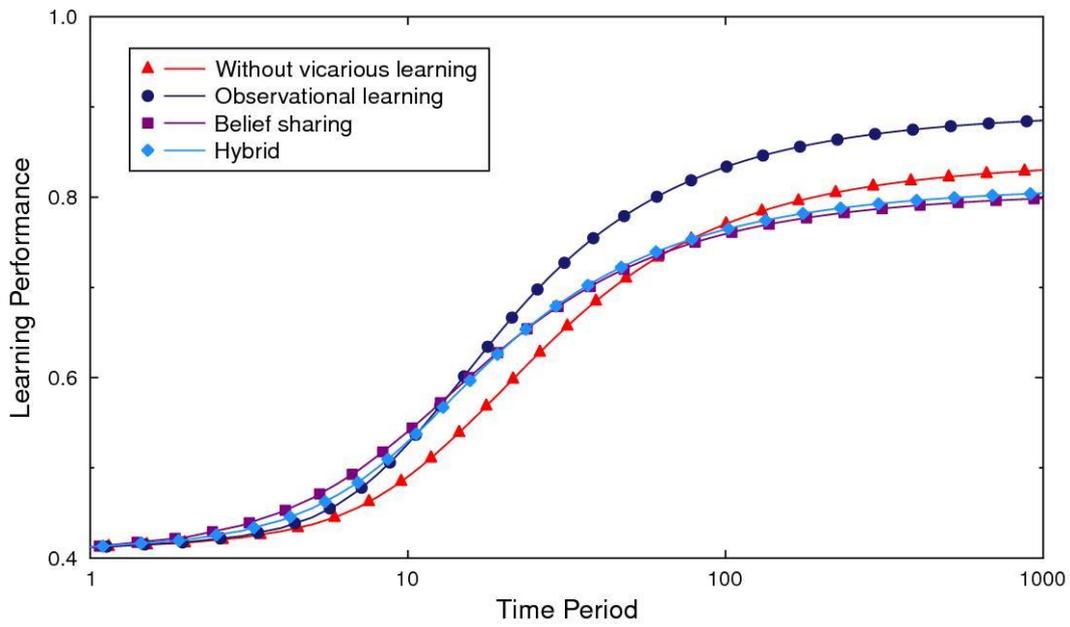

**Figure 3. Intertemporal behaviors of the learning system w.r.t. vicarious learning mode ($\phi = 0.5, \Pi_{max} = 1, \alpha = 0.8, \varepsilon = 0.1$)**

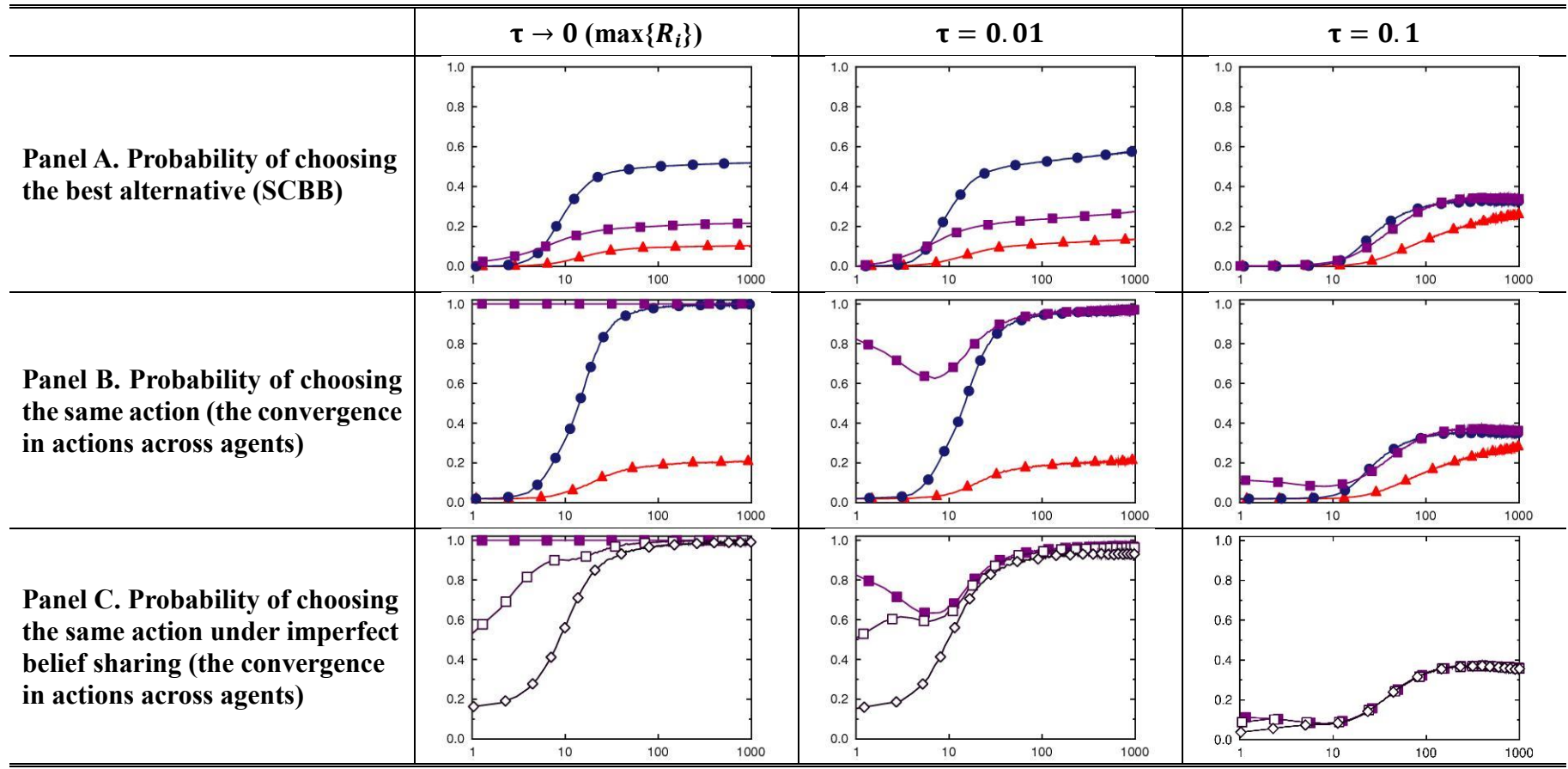

* The number in the parenthesis represents the weight on other's belief

**Figure 5. Vicarious learning without endogenous information ($\tau \to 0$ ($\max\{R_i\}$), $\phi = 0.5, \Pi_{max} = 1, \alpha = 0.8, \varepsilon = 0.1, T = 1,000$)**

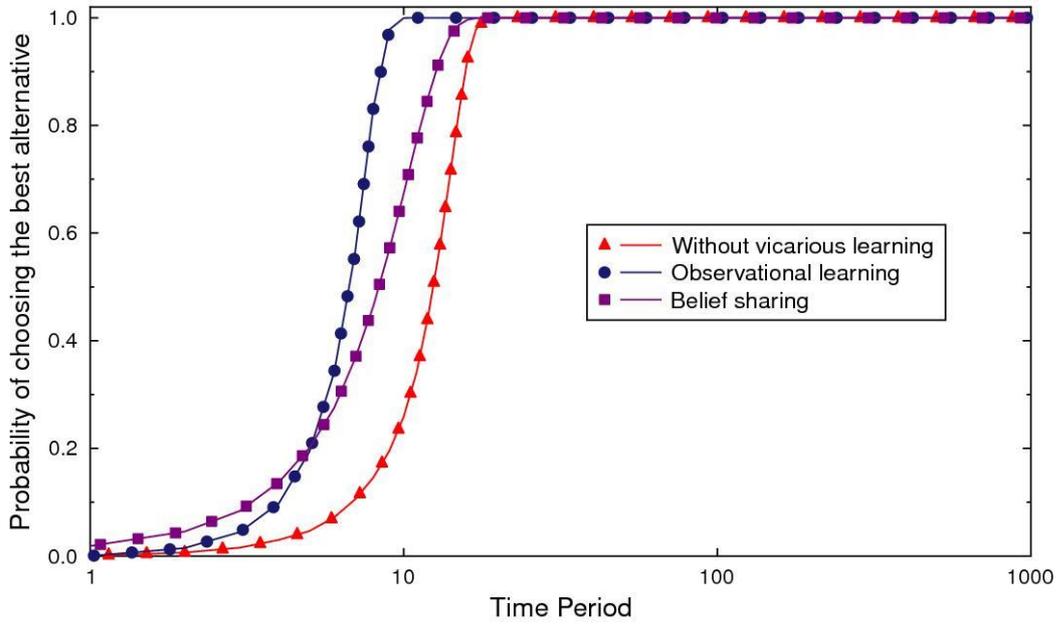

**Figure 6. Learning performance of the system with incomplete observation**

**(a) Observing outcome only**[3]

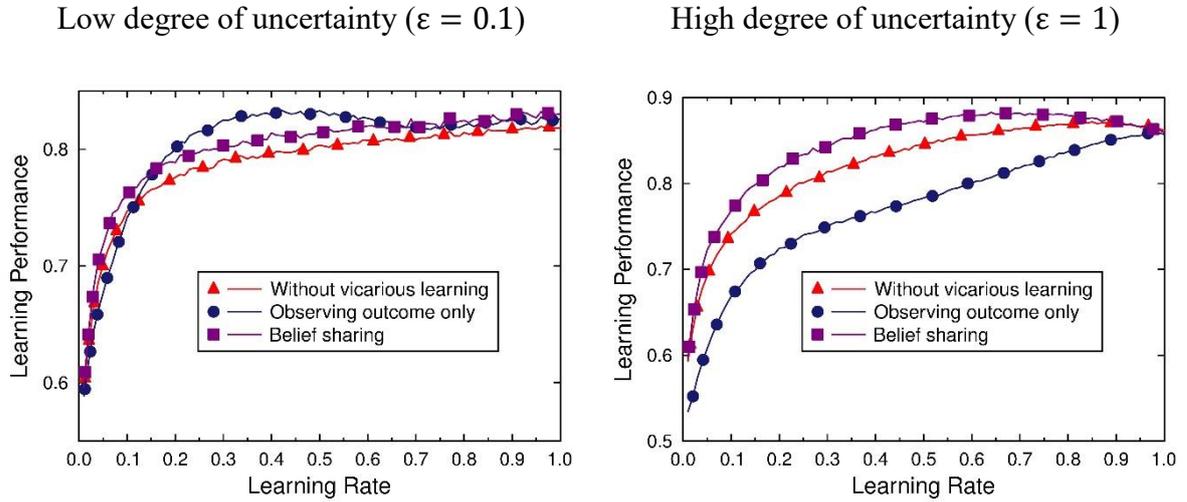

Low degree of uncertainty ($\varepsilon = 0.1$)　　　　High degree of uncertainty ($\varepsilon = 1$)

**(b) Observing action only**[4]

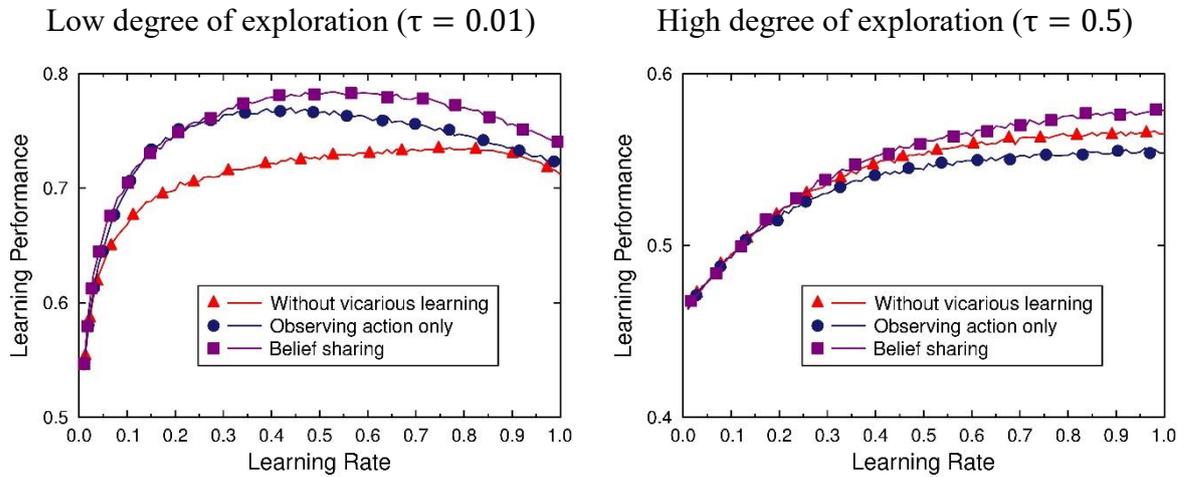

Low degree of exploration ($\tau = 0.01$)　　　　High degree of exploration ($\tau = 0.5$)

---

[3] For this analysis, we set our parameters as follows: $\tau = 0.01$, $m = 5$, $\alpha = 0.8$, $\Pi_{max} = 1$, $T = 50$, $\tau_H = 0.1$, $\tau_L = 0.01$, and $c = 1.5$.

[4] For this analysis, we set our parameters as follows: $m = 10$, $\alpha = 0.8$, $\Pi_{max} = 1$, $T = 50$, $\tau = 0.01$, and $\varepsilon = 1$.

**Figure 7. The effect of search space ($\Pi_{max} = 1, \alpha = 0.8, \varepsilon = 1, T = 50,$ and $\tau = 0.01$)**

(a) Narrow search space ($m = 10$)

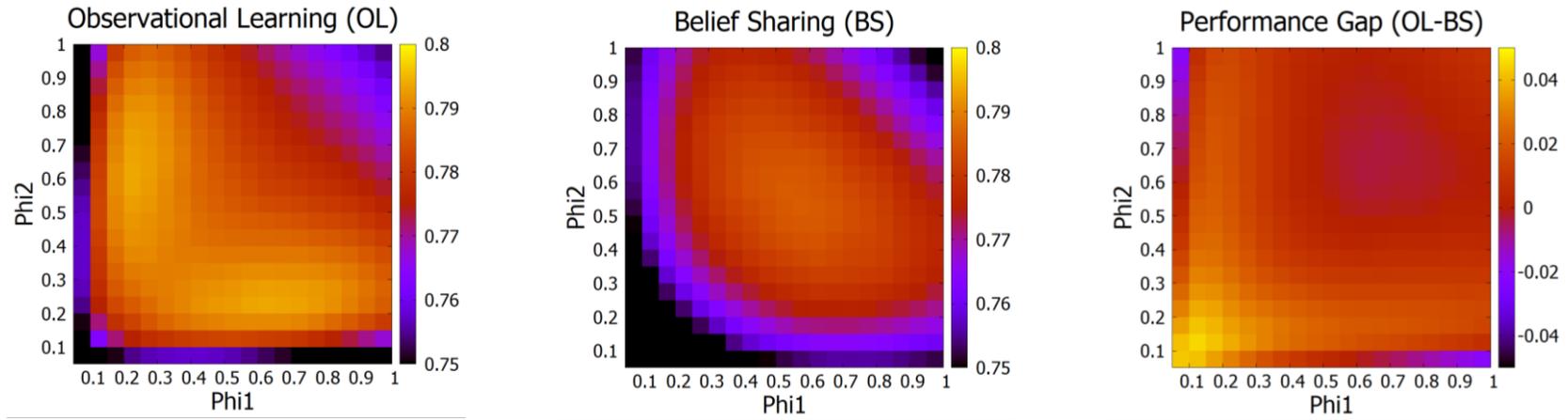

(b) Broad search space ($m = 50$)

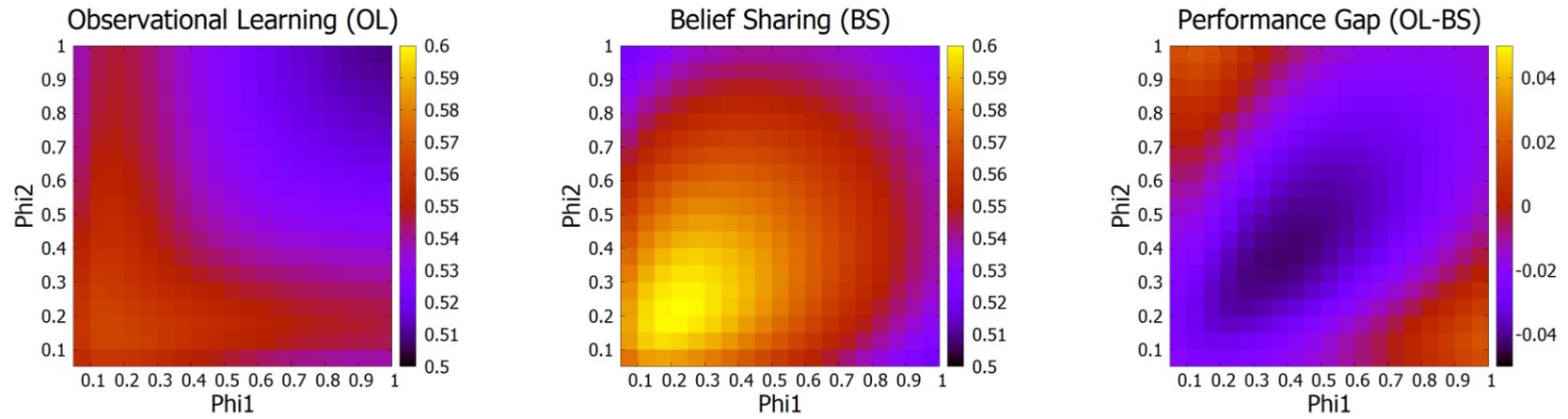

**Figure 8. The effect of variance in payoff distribution (i.e. spikiness)** ($m = 50$, $\varepsilon = 1$, $T = 50$, and $\tau = 0.01$)

(a) Low variance ($\Pi_{max} = 1$ and $\alpha = 0.8$)

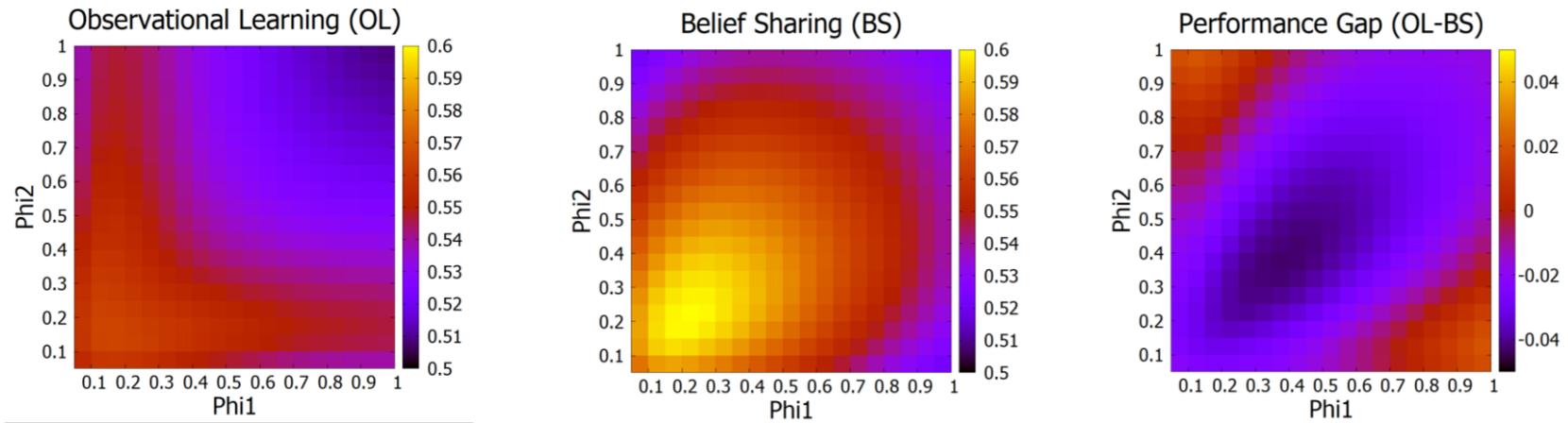

(b) High variance ($\Pi_{max} = 1.5$ and $\alpha = 0.8$)

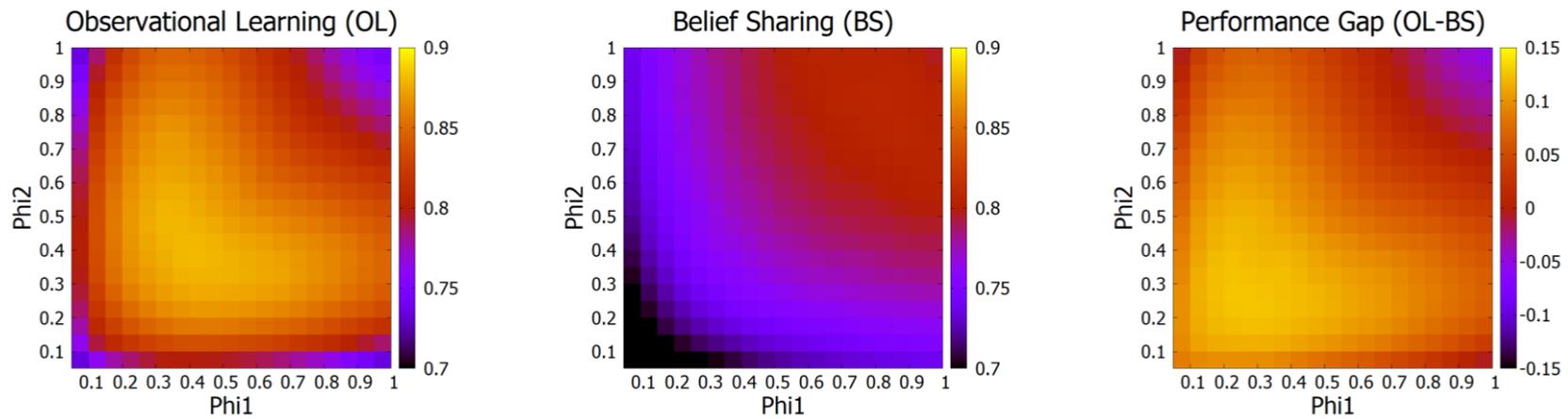

**Figure 9. The effect of learning periods** ($m = 50, \Pi_{max} = 1, \alpha = 0.8, \varepsilon = 1, \tau = 0.01$)

(a) Limited learning period ($T = 50$)

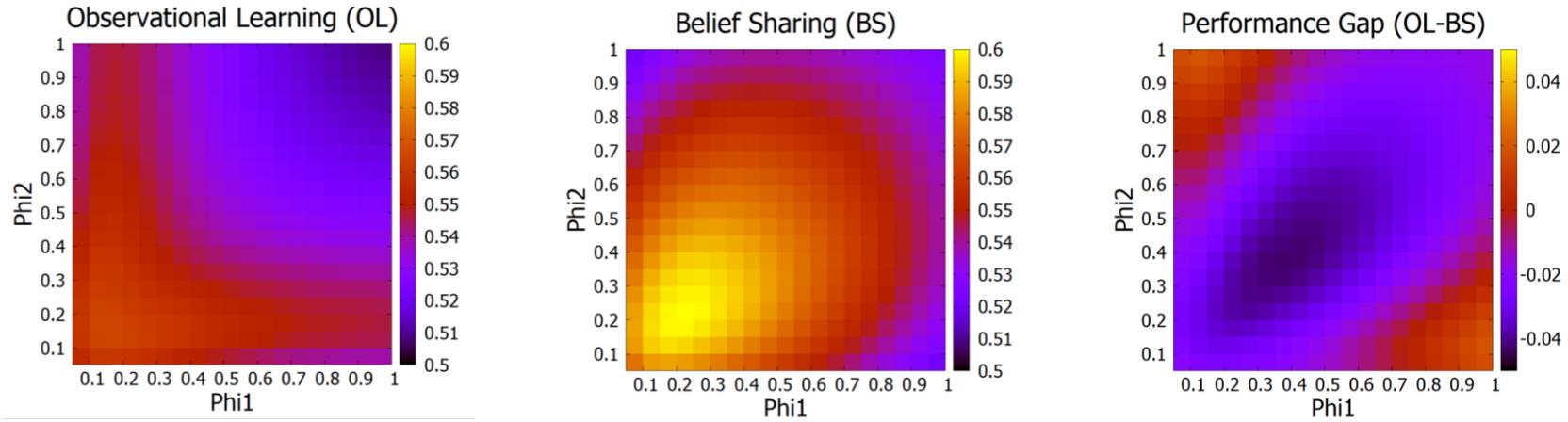

(b) Sufficient learning period ($T = 1,000$)

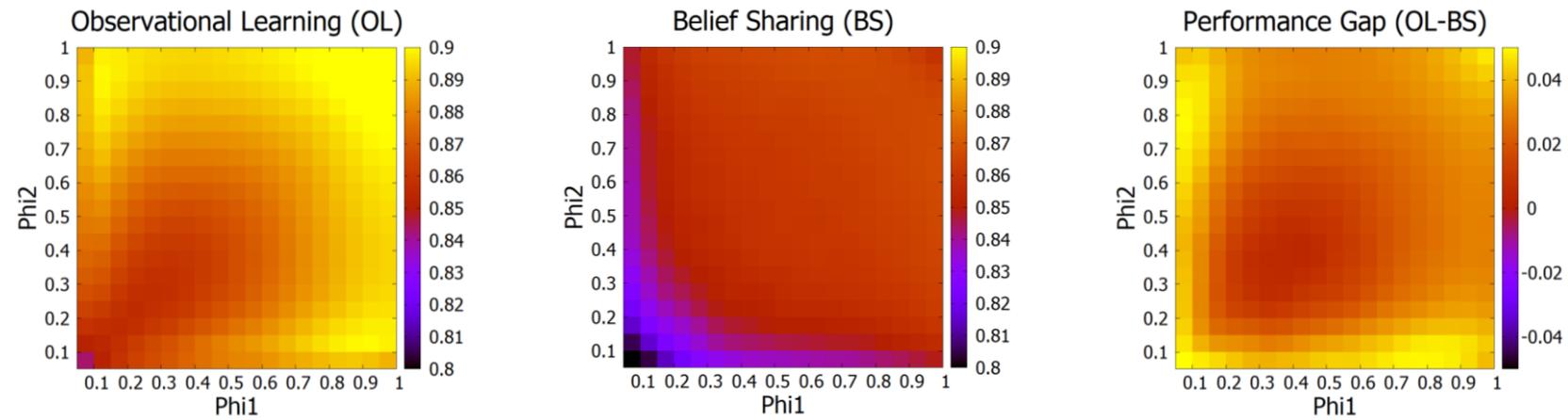

# APPENDIX

**Appendix A. The convergence of action within agent (switching behaviour)**

**Figure A1. The probability of switching action**

**(a) $\tau \to 0$ ($\max\{R_i\}$)**

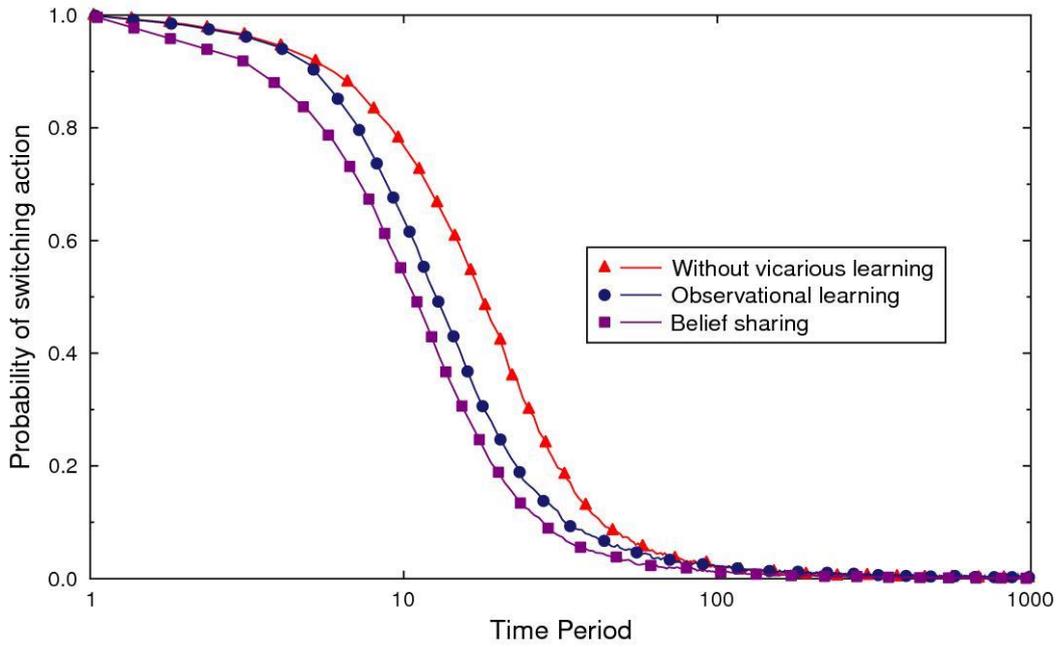

**(b) $\tau = 0.01$**

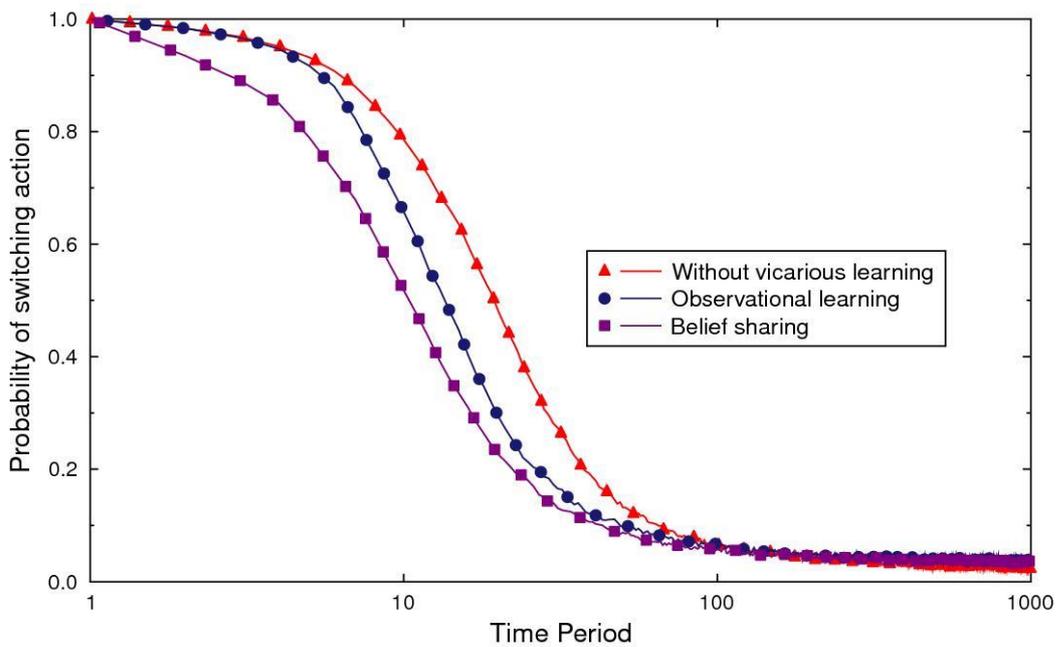

**(c) τ = 0.1**

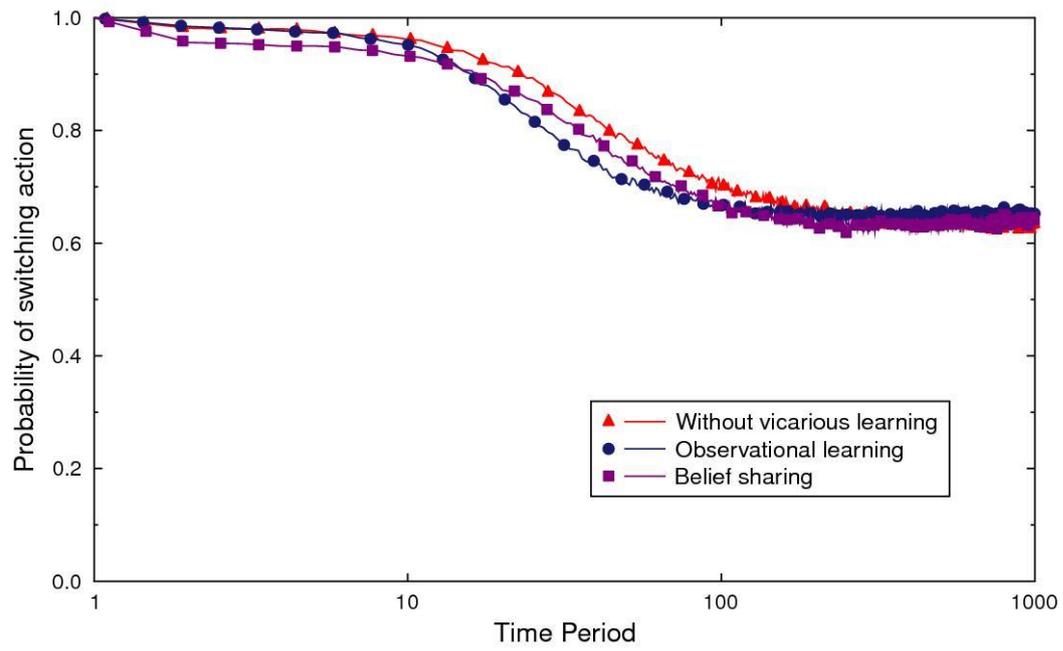

**Appendix B. Does asymmetry improve learning performance?**

In learning literature, researchers have emphasized the benefit of having asymmetric agents in the system. For example, March (1991) argues that agents with heterogeneous socialization rates enable an organization to balance exploitation and exploration. In addition, Lave and March (1993) contends that asymmetry in learning rates can redress the coordination problem. Following this tradition, we explore whether asymmetry in learning rates improves the learning performance of the collaborative learning system as well. In particular, we focus on how asymmetry balances the impacts of prior and uncertainty on self-confirming biased beliefs.

Figure A2 represents the learning performance with respect to the different combinations of learning rates. We find that asymmetry in learning rates improves the learning performance of the system with observational learning (Figure A2a). This is because, in observational learning, slow and fast learners can complement each other. To be specific, a slow learner (i.e., less sensitive to the recent outcome) provides stability in sampling, thereby reducing the impact of uncertainty on self-confirming biased beliefs. On the other hand, a fast learner enables the system to rapidly deviate from false positive beliefs, thereby reducing the impact of prior on self-confirming biased beliefs.

However, in the system with belief exchange (Figure A2b), asymmetric learning rates reduces the learning performance. This is because, in the system with belief sharing, the asymmetric agents cannot balance the contradictory effects of uncertainty since the agents build a shared belief system. Moreover, when the agents with different learning rates construct a shared belief system, information obtained by a fast learner outweighs that obtained by a slow learner. Considering that the quality of information is identical across the agents, assigning different weights to information will increase variance in the inference. This implies that a mode of vicarious learning determines the system's ability to redress self-confirming biased beliefs by asymmetry. In other words, in the collaborative learning system, the benefit of asymmetric agents can be fully realized only when their belief systems are isolated.

**Figure A2. Learning performance of the system with asymmetric agents**

(a) Observational learning  (b) Belief sharing

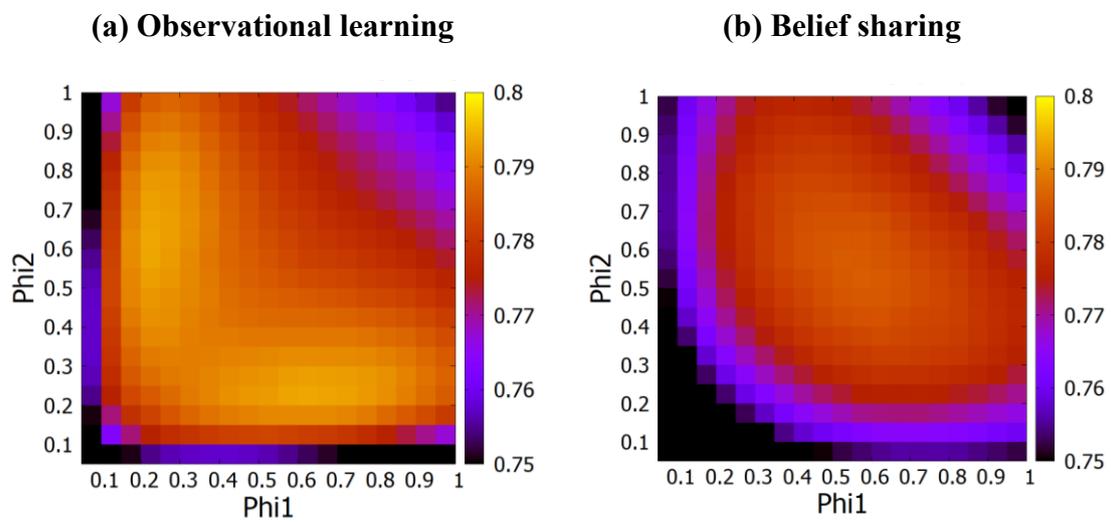

**Appendix C. How does vicarious learning affect the search scope?**

In this section, we examine how vicarious learning affects the search scope of the agent and the system. We define the search scope as the number of sampled alternatives. For instance, when the agent tried five different alternatives in a given period (i.e., T = 100), the search scope will become 5. On the other hand, the search scope of the system is defined as a union of two agents' search scope. For example, when the agent 1 tried three alternatives (1, 3, and 5) and the agent 2 tried three alternatives (1, 2, and 5), the search scope of the system will be 4 as four alternative was sampled by the system (i.e., 1, 2, 3, and 5). Figure A3 shows that the system with observational learning tends to have a broader search scope than belief sharing. Interestingly, we find that vicarious learning curtails the search scope at both levels. This implies that vicarious learning improves the effectiveness of search given that it also increases the probability to find the best alternative (Figure 3).

**Figure A3. Vicarious learning and the search scope of the agent and the system[5]**

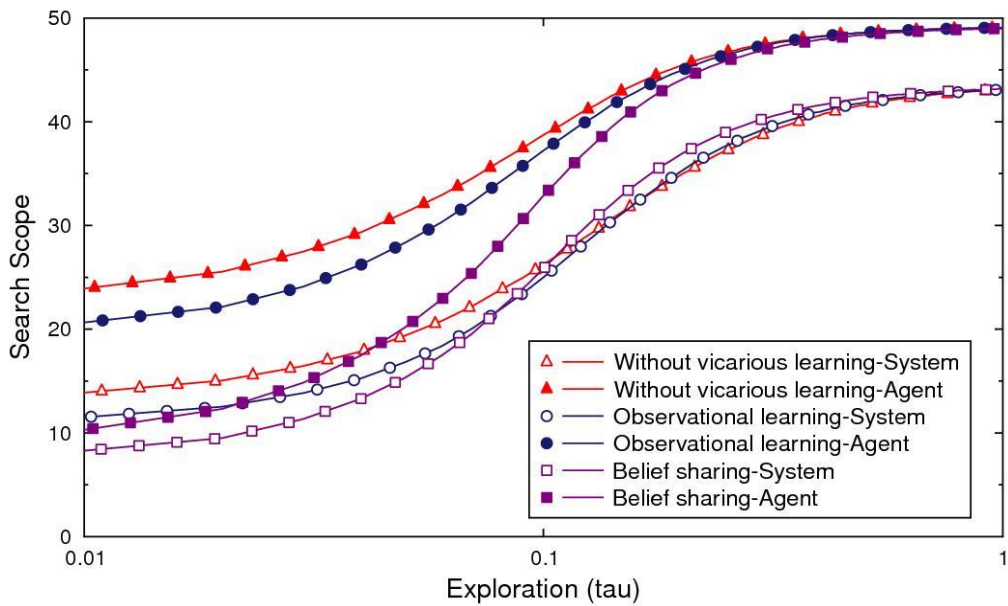

---

[5] For this analysis, we set our parameters as follows: $\phi = 0.5$, $m = 50$, $\alpha = 0.8$, $\varepsilon = 1$, $T = 100$.

**Appendix D. Different updating rules and self-confirming biased beliefs**

**Figure A4. Self-confirming biased beliefs with respect to updating rules ($\tau = 0.01$)**

**(a) EWA – EWA**

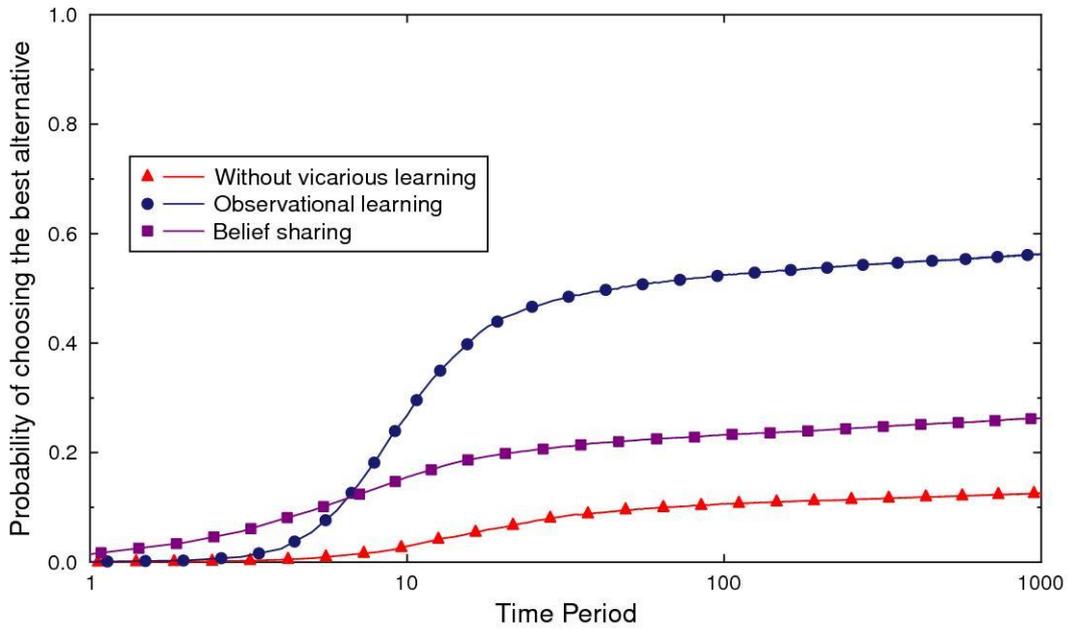

**(b) Averaging – EWA**

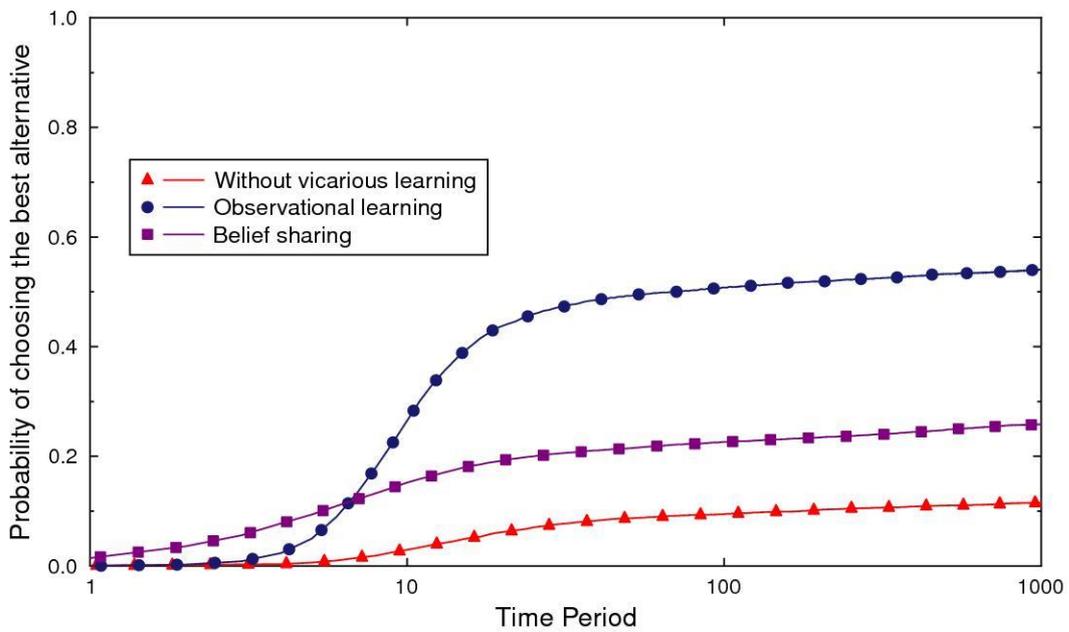

**(c) Averaging – Averaging**

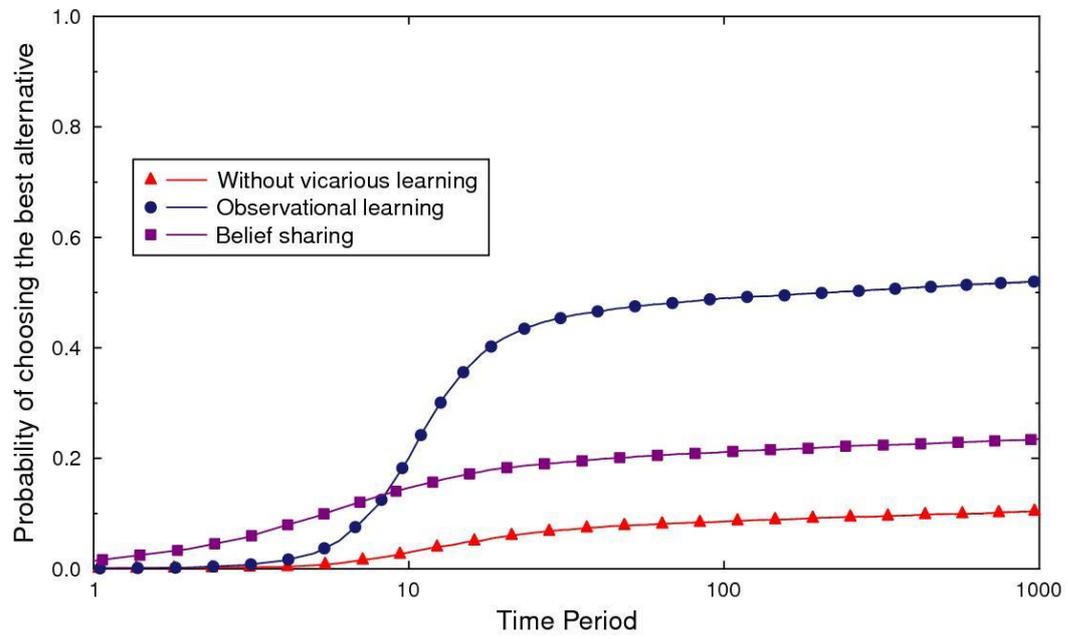

**Appendix E. Unbiased prior and feedback without noise**

**Figure A5. Self-confirming biased beliefs with EWA**

**(a) τ → 0**

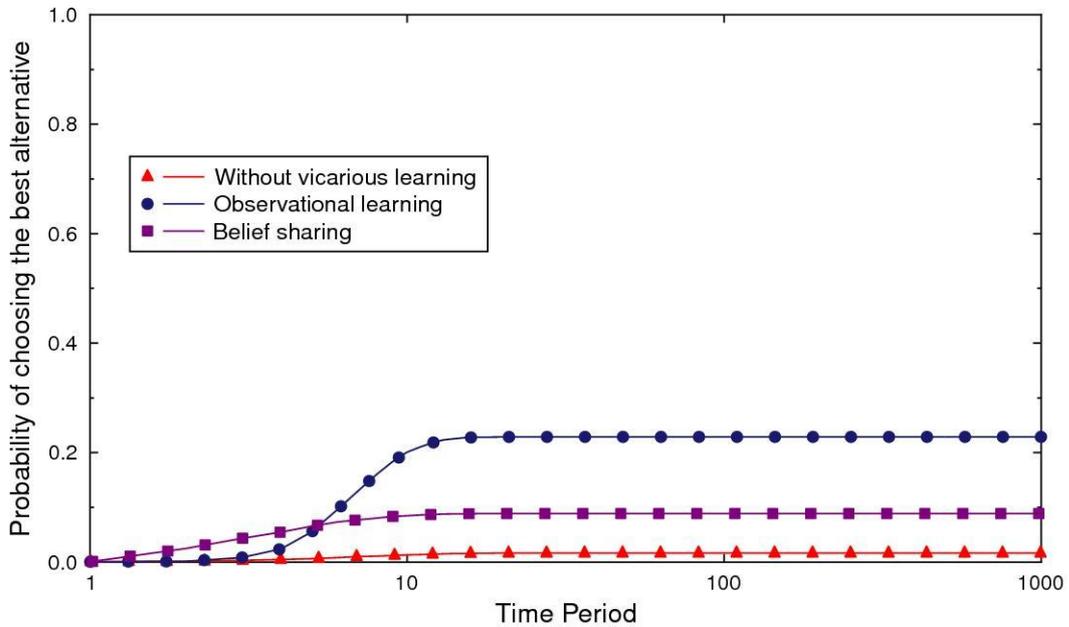

**(b) τ = 0.01**

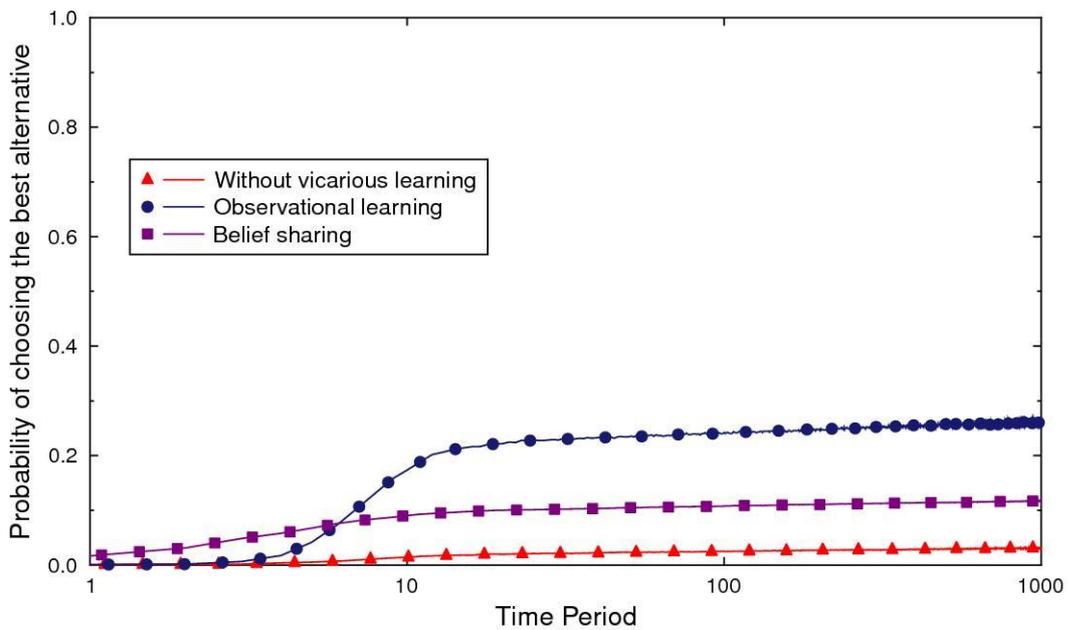

**(c) τ = 0.1**

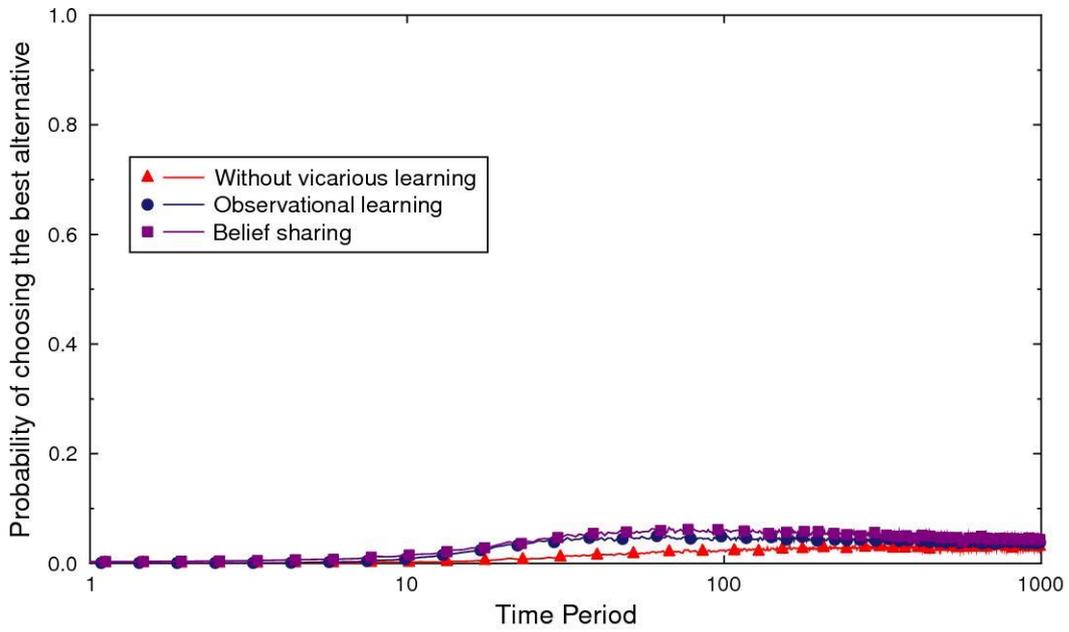

**Figure A6. Self-confirming biased beliefs with Bayesian updating (averaging)**

**(a) τ → 0**

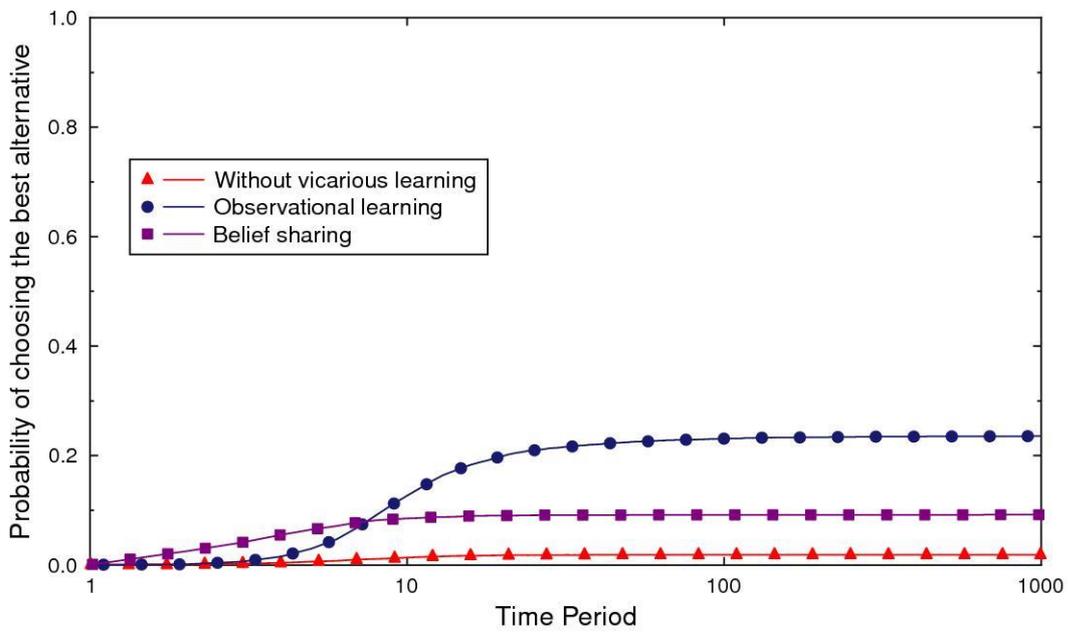

**(b) $\tau = 0.01$**

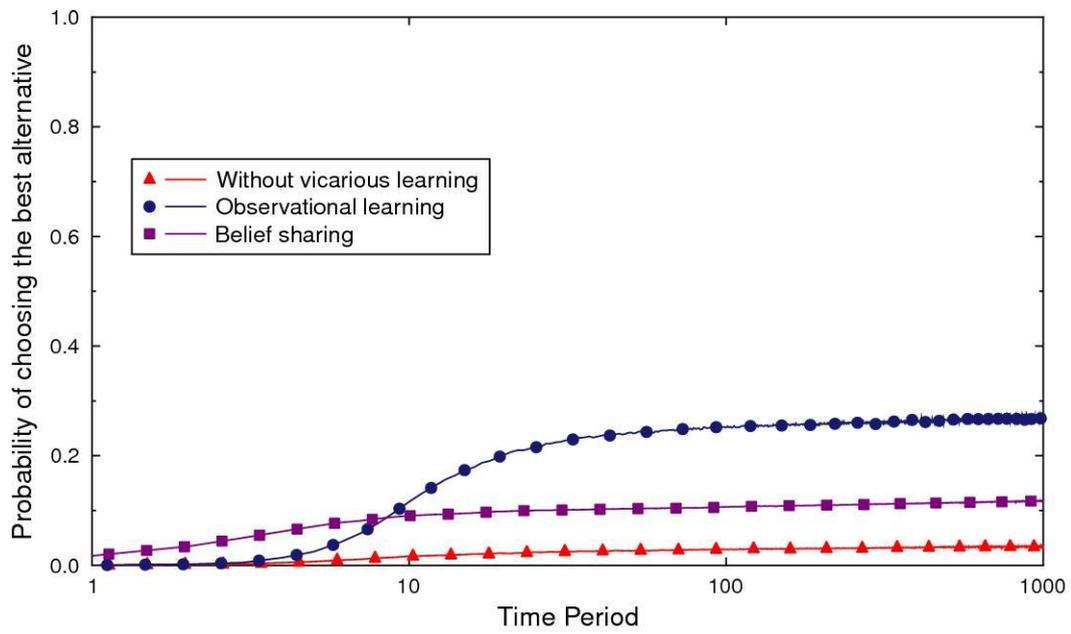

**(c) $\tau = 0.1$**

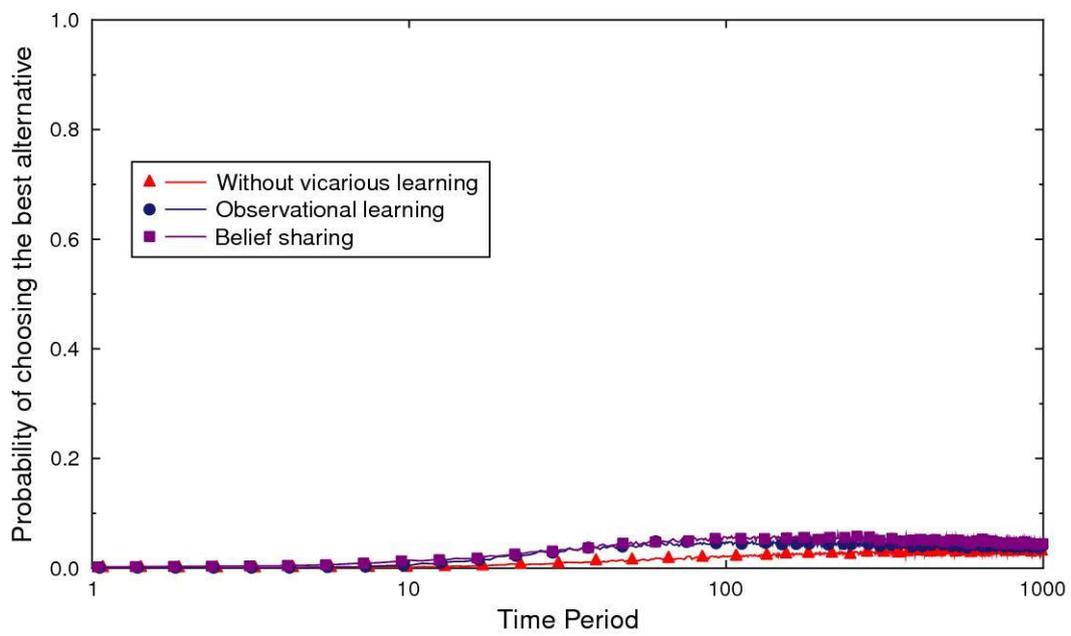

**Appendix F. Larger systems with vicarious learning**

**Figure A5a. Random network (ER model with $n = 100, p = 0.02$)[6]**

*When the self-confirming biased beliefs is critical ($\Pi_{max} = 2$)*

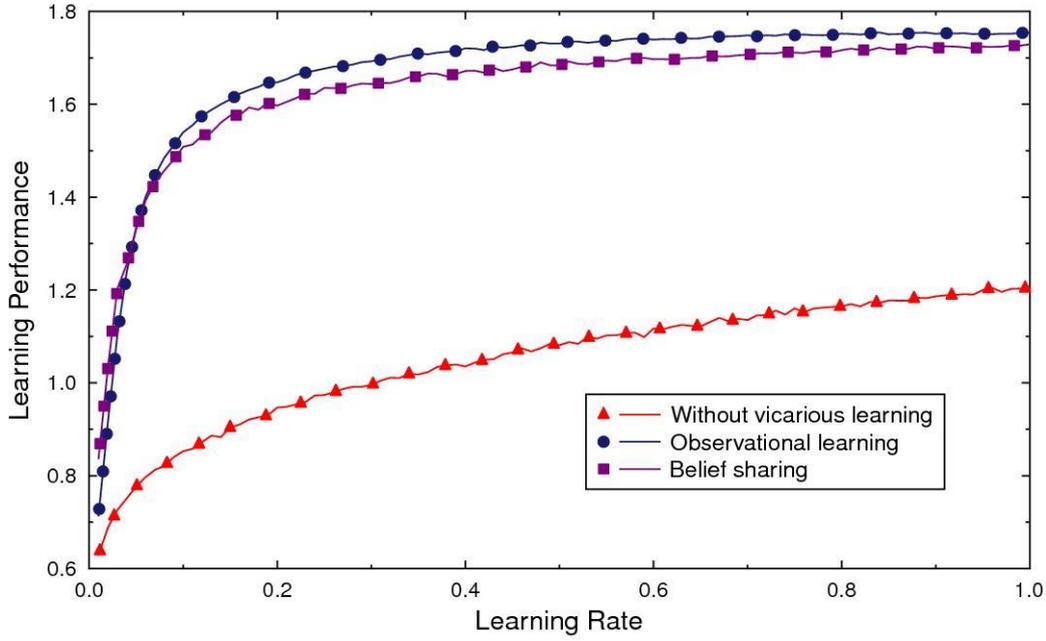

*When the self-confirming biased beliefs is less critical ($\Pi_{max} = 1$)*

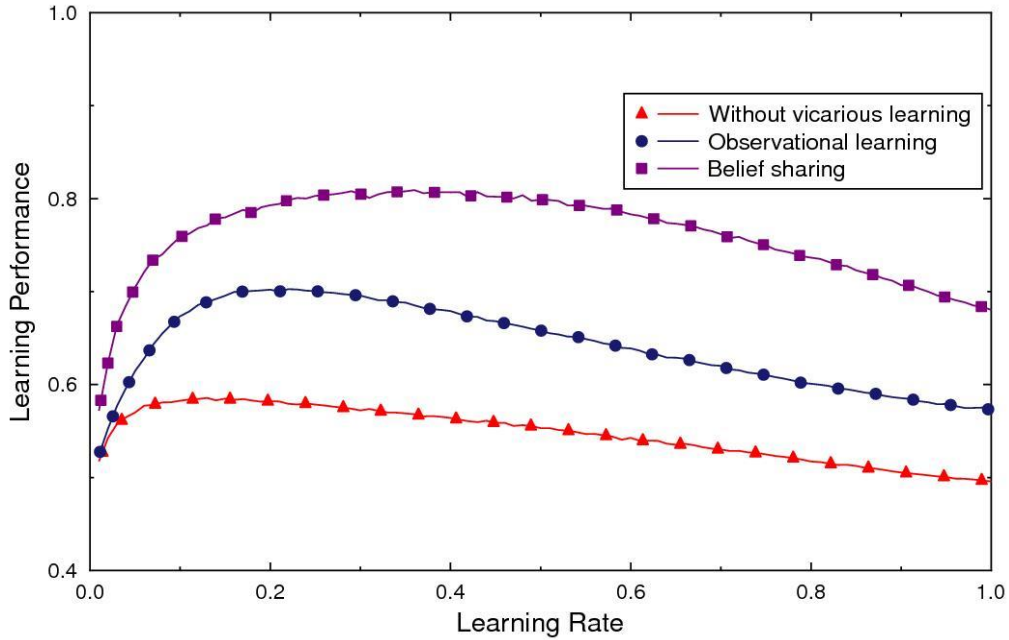

---

[6] For this analysis, we set our parameters as follows: $\tau = 0.01$, $m = 50$, $\alpha = 0.8, \varepsilon = 1$, $T = 100$.

**Figure A5b. 2D lattice with Von Neumann neighbourhood (5X5)**

*When the self-confirming biased beliefs is critical ($\Pi_{max} = 2$)*

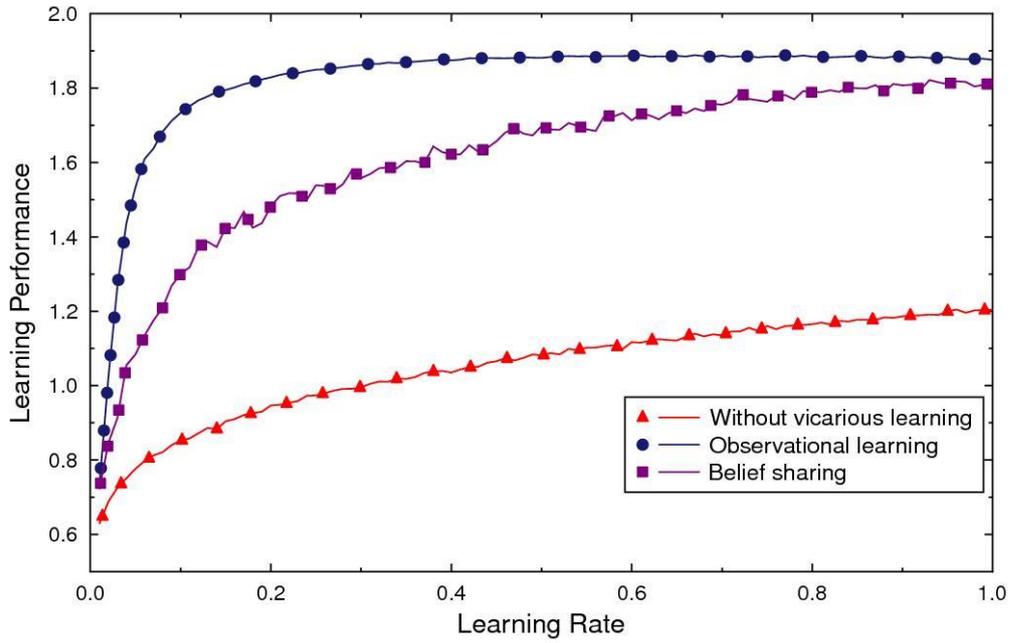

*When the self-confirming biased beliefs is less critical ($\Pi_{max} = 1$)*

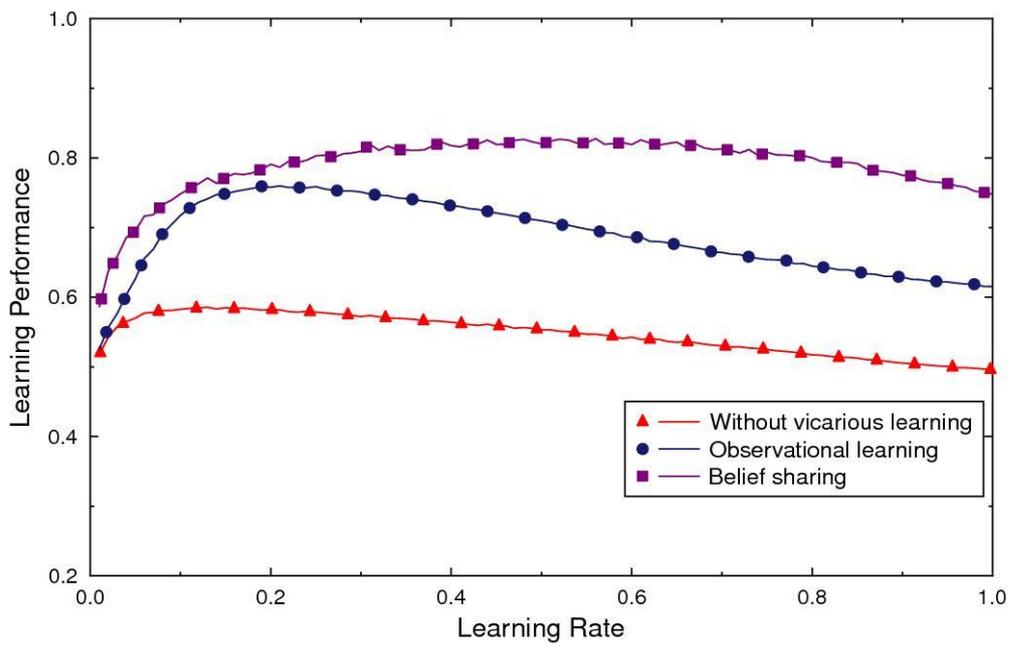

**Appendix G. Incomplete belief sharing and self-confirming biased beliefs**

**Figure A8. Self-confirming biased beliefs with respect to belief sharing**

**(a) Sharing a limited number of dimensions ($\tau \to 0$)**

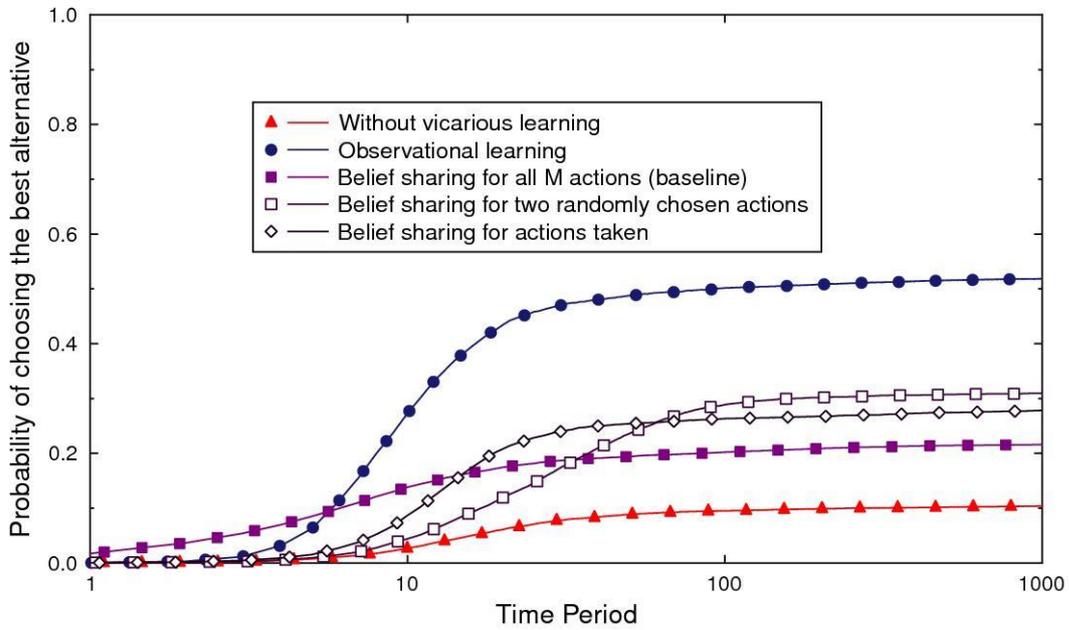

**(b) Belief sharing with different frequencies ($\tau = 0.01$)**

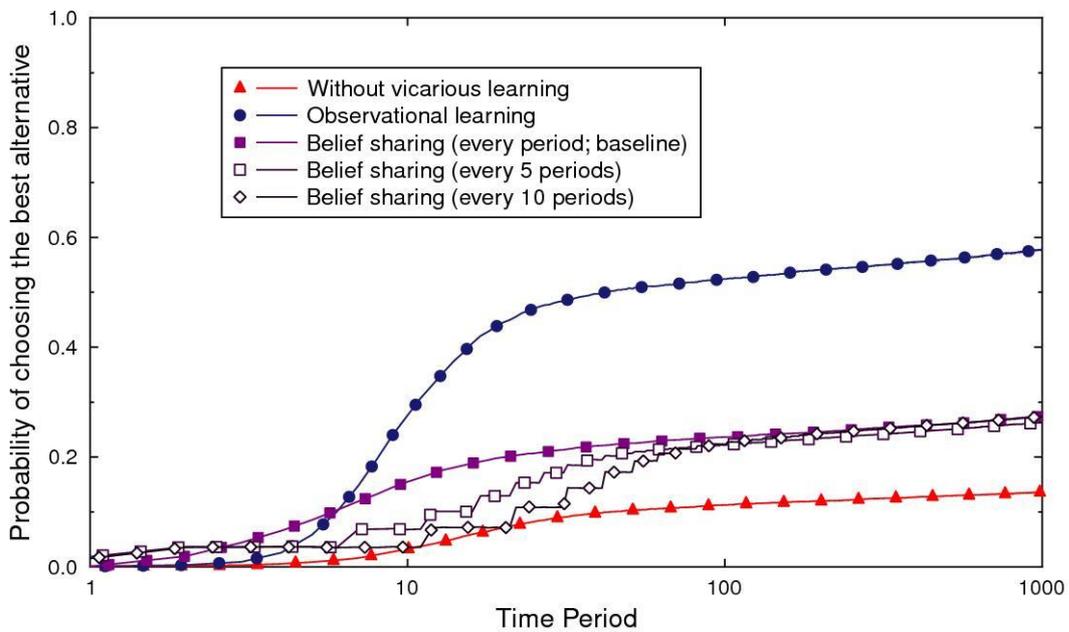